\documentclass{aa501}
\usepackage{graphicx}

\newcommand{\kmprs}  {\mbox{\rm km\,s$^{-1}$}}

\newcommand{\feh} {\mbox{\rm [Fe/H]}}
\newcommand{\ofe} {\mbox{\rm [O/Fe]}}

\newcommand{\alphafe} {\mbox{\rm [$\alpha$/Fe]}}

\newcommand{\teff}  {\mbox{$T_{\rm eff}$}}
\newcommand{\logteff} {\mbox{${\rm log}\,T_{\rm eff}$}}
\newcommand{\logg}  {\mbox{{\rm log}\,$g$}}
\newcommand{\LiI} {\ion{Li}{i}}
\newcommand{\OI} {\ion{O}{i}} 
\newcommand{\OII} {\ion{O}{ii}} 
\newcommand{\forOI} {\mbox{\rm [\ion{O}{i}]}}
\newcommand{\FeI} {\ion{Fe}{i}}
\newcommand{\FeII} {\ion{Fe}{ii}}
\newcommand{\ScII} {\ion{Sc}{ii}}
\newcommand{\NiI} {\ion{Ni}{i}}
\newcommand{\SiI} {\ion{Si}{i}}

\newcommand{\BeII} {\ion{Be}{ii}}
\newcommand{\Mv} {\mbox{$M_V$}}

\newcommand{\lirat} {\element[][6]{Li}/\element[][7]{Li}}

\begin{document}

\title{O/Fe in metal-poor main sequence and subgiant stars
\thanks{Based on observations collected at the European
Southern Observatory, Chile (ESO No. 65.L-0131, 65.L-0507,
and 67.D-0439)}}

%\subtitle{}

\author{P.E.~Nissen \inst{1} \and F.~Primas \inst{2} 
\and M.~Asplund \inst{3,4} \and D.L. Lambert \inst{5}}

\offprints{P.E.~Nissen}

\institute{
Department of Physics and Astronomy, University of Aarhus, DK--8000
Aarhus C, Denmark\\
\email{pen@ifa.au.dk}
\and European Southern Observatory, Karl-Schwarzschild Str. 2,
D--85748 Garching b. M\"{u}nchen, Germany\\
\email{fprimas@eso.org}
\and Uppsala Astronomical Observatory, Box 515, SE--751 20, Sweden\\
\and
Present address: Research School of Astronomy and Astrophysics,
Australian National University, Mount Stromlo Observatory,
Cotter Road, Weston, ACT 2611, Australia\\
\email{martin@mso.anu.edu.au}
\and Department of Astronomy, University of Texas, Austin, TX 78712--1083\\
\email{dll@anchor.as.utexas.edu}
}

\date{Received ..... / Accepted ......}

\abstract{A study of the O/Fe ratio in metal-poor main sequence and subgiant 
stars is presented using the \forOI\,6300 \AA\ line, the \OI\ 7774\,\AA\ 
triplet, and a selection of weak \FeII\ lines observed on high-resolution
spectra acquired with the VLT UVES spectrograph. The \forOI\ line is 
detected in the spectra of 18 stars with $-0.5 < \feh -2.4$, and the triplet 
is observed for 15 stars with \feh\ ranging from $-1.0$ to $-2.7$. The 
abundance analysis was made first using standard model atmospheres taking 
into account non-LTE effects on the triplet: the \forOI\ line and the 
triplet give consistent results with [O/Fe] increasing quasi-linearly
with decreasing [Fe/H] reaching [O/Fe] $\simeq$ $+0.7$ at [Fe/H] = $-2.5$.
This trend is in reasonable agreement with other results for [O/Fe] in 
metal-poor dwarfs obtained using standard atmospheres and both ultraviolet 
and infrared OH lines. There is also broad agreement with published results 
for [O/Fe] for giants obtained using standard model atmospheres and the 
\forOI\ line, and the OH infrared lines, but  the \OI\ lines give higher 
[O/Fe] values which may, however,  fall into place when non-LTE effects are 
considered. When  hydrodynamical model atmospheres representing stellar
granulation in dwarf and subgiant stars replace standard models, the [O/Fe] 
from the \forOI\ and \FeII\ lines is decreased by an amount which increases 
with decreasing [Fe/H]. These 3D effects on [O/Fe] is compounded by the 
opposite behaviour of the \forOI\ (continuous opacity effect) and \FeII\ 
lines (excitation effect). The [O/Fe] vs [Fe/H] relation remains 
quasi-linear extending to [O/Fe]  $\simeq$ +0.5 at [Fe/H] = $-2.5$, but with 
a tendency of a plateau with [O/Fe] $\simeq$ +0.3 for $-2.0 <$ [Fe/H] 
$< -1.0$, and a hint of cosmic scatter in [O/Fe] at $\feh \simeq -1.0$.
Use of the hydrodynamical models disturbs the broad agreement between the 
oxygen abundances from the \forOI , \OI\, and OH lines, but 3D non-LTE 
effects may serve to erase these differences. The [O/Fe] values from the 
\forOI\ line and the hydrodynamical model atmospheres for dwarfs and 
subgiant stars are lower than the values for giants using standard model 
atmospheres and the \forOI\, and \OI\ lines.
\keywords{Stars: abundances -- Stars: atmospheres -- Stars: fundamental
parameters -- Galaxy: evolution -- }}

\maketitle

\section{Introduction}

The oxygen abundance of metal-poor stars is intimately linked to
several outstanding questions in astrophysics. Questions
of current interest include the following three: How old are the oldest
(i.e., the most metal-poor) stars? Can the observed chemical evolution
of oxygen in  the
early Galaxy be interpreted using current predictions for the yields
from metal-poor massive stars exploding as supernovae? 
What conclusions may be drawn from the abundance ratios of
lithium, beryllium, and boron relative to oxygen about the synthesis of
these three light elements by collisions involving high energy particles
in  interstellar circumstellar gas? For these and other reasons,
accurate determinations of the growth of the oxygen abundance of 
the Galaxy are desirable. 

Oxygen, the most abundant element after hydrogen and helium, affects
the opacity and, hence, the structure and evolution of low mass stars.  
Globular clusters are considered to contain some of the oldest stars in the
Galaxy. Oxygen affects the shape of a cluster color-magnitude
diagram, and, hence, a cluster's age as it is determined from the fit of
a theoretical to the observed diagram. VandenBerg \& Bell (\cite{vandenberg01}) estimate
that `each increase of 0.3 dex in [O/Fe] implies a 1 Gyr age reduction
at a fixed turnoff luminosity'. In the case of the most metal-poor stars,
current estimates of the [O/Fe] differ by as much as 0.7 dex or ages
differ by 2 to 3 Gyr, a not insignificant difference for
cosmologists.

Compositions of metal-poor unevolved stars are presumed to reflect
the composition of the gas from which they formed. In the case of the
most metal-poor stars, that gas may have been primordial gas mixed
with ejecta from just one or two Type II supernovae. Hence, the
predicted compositions of the ejecta may be tested against observed
compositions. Predictions appeared not to have a problem in
matching observations that suggested [O/Fe] $\simeq$ 0.4, as well as
[$\alpha$/Fe] $\simeq$ 0.4, where $\alpha$ denotes Mg, Si, and Ca.
 With some observers now reporting much higher
oxygen abundances, [O/Fe] $\simeq$ 1 for very metal-poor stars,
serious challenges may face the modelers of the Type II
supernova (see particularly Goswami \& Prantzos  \cite{goswami00}).
Recent claims that  hypernovae,
 i.e. supernovae with very 
high explosion energies, at low metallicities could account for 
the  high values of  [O/Fe]  (Israelian et al.  \cite{israelian01}) 
appear not to be
supported by the most recent theoretical investigations of such events.
On the contrary, detailed nucleosynthetic calculations reveal significantly
{\em lower}  O/Fe ratios than in normal core-collapse supernovae
(Nakamura et al. \cite{nakamura01}).

Beryllium is identified as solely a product of spallation of
nuclei in
interstellar or circumstellar gas traversed by energetic particles:
for example, the process $p$ + O $\rightarrow$ splits $^9$Be nuclei from the
oxygen $^{16}$O nucleus.
Spallation also makes a contribution to the synthesis of lithium and
boron. Under quite
simple but reasonable assumptions concerning the nucleosynthesis of
oxygen and the origin of the energetic particles (here referred to
as cosmic rays, or CR for short), if beryllium synthesis is
controlled by a CR proton (or $\alpha$-particle) colliding with an
interstellar oxygen nucleus, one expects the beryllium abundance
to grow as the square of the oxygen abundance. If the roles are
reversed and  a cosmic ray oxygen nucleus collides with an interstellar
proton (or $\alpha$-particle), the beryllium and oxygen abundances
are linearly related. In the present interstellar medium, this latter
process is a minor  contributor to beryllium synthesis because
the fragments of the spallation of the oxygen nuclei have high energies
and are themselves spallated as they are thermalized. In the former
process, the fragments have much lower energies and  a
higher survival rate in thermalization. These relative roles
may be reversed in the early Galaxy when the interstellar oxygen
abundance was much lower than in the interstellar medium at
present. By determining the Be/O ratio for metal-poor (and metal-rich)
stars, it is possible to gain insight into the spallation processes. 
Uncertainty about the oxygen abundance necessarily compromises
this  exercise. Substitution of iron for oxygen is unsatisfactory
because very little beryllium is produced from iron, and also the
predicted oxygen to iron ratio in ejecta of  Type II supernovae
is uncertain.

Spectra of metal-poor stars offer several potential indicators of
the oxygen abundance: ultraviolet OH lines, the [O\,{\sc i}]
6300 \AA\ and 6363 \AA\ lines, the O\,{\sc i} triplet at 7774 \AA,
and infrared vibration-rotation lines from the OH molecule's
ground state. A selection of these lines is available for
both metal-poor dwarfs and giants. A problem of great current
concern is that the various indicators may give different results.
An abbreviated discussion suffices here, and a fuller discussion
is postponed to Section 5. 

The problem became strikingly evident with the publication of
analyses of the ultraviolet OH lines in dwarfs 
(Israelian et al. \cite{israelian98}; Boesgaard et al.
\cite{boesgaard99}). These independent but quite similar
analyses showed that  [O/Fe] rose linearly
with [Fe/H] from zero at [Fe/H] = 0 to about 1.2 at [Fe/H] = $-3.0$,
about the metallicity of the most metal-poor stars studied.
Accounting for non-LTE effects on the reference \FeI\ lines diminishes
the slope somewhat to $-0.33$ but still with a monotonic trend
(Israelian et al. \cite{israelian01}).
The linear increase to lower metallicities anticipated by Abia \&
Rebolo (\cite{abia89}) from the O\,{\sc i} lines was confirmed by some
but not all analyses of these  lines.
A quite different relation for [O/Fe] versus [Fe/H] had been derived
prior to the OH analyses from the [O\,{\sc i}] lines in giants:
[O/Fe] $\simeq$ 0.4 independent of [Fe/H] for giants
with [Fe/H] $< -1$ (Barbuy \cite{barbuy88}). At [Fe/H] = $-3$,
the OH in dwarfs and the [O\,{\sc i}] in giants results for
[O/Fe] differ by about 0.6 dex.

A strong {\it a priori} claim may be advanced  that the
[O\,{\sc i}]-based analysis is  the least sensitive to systematic
errors: the lines are formed in or very close to local thermodynamic
equilibrium (LTE); essentially all of the oxygen atoms in the
dwarf and giant atmospheres are in the ground
configuration providing the lower and the upper level of the
forbidden lines. Even the sensitivity of the [O\,{\sc i}]-based
results to the assumed surface gravity is effectively mitigated
by determining the iron abundance from Fe\,{\sc ii} lines. This
pairing also ensures the non-LTE effects affecting the iron abundance
are minimal; stronger non-LTE effects are predicted for Fe\,{\sc i}
lines (e.g. Th\'{e}venin \& Idiart \cite{thevenin99}). 
Dependence of [O/Fe] on effective temperature is also quite 
insignificant when
the combination of [O\,{\sc i}] and Fe\,{\sc ii} lines is used, whereas
the dependence on the atmospheric structure in general cannot be neglected,
as will be discussed later, but at least it is smaller than in
the case where OH lines are used.

By sharp contrast, the number of OH molecules (and of O\,{\sc i} atoms in
the lower state of the 7774 \AA\ transitions) is a very small fraction of the
total number of oxygen atoms, a fraction that is highly sensitive to the
adopted values of effective temperature and surface gravity. Furthermore,
LTE is far from guaranteed for either the OH UV lines 
(Hinkle \& Lambert \cite{hinkle75};
 Asplund \& Garc\'{\i}a P{\'e}rez \cite{asplund01}) or the
O\,{\sc i} 7774 \AA\ lines (Kiselman \cite{kiselman01} and references therein).

The view has been expressed that the [O\,{\sc i}]-based
results for the giants are in error on two accounts
 (Takada-Hidai et al. \cite{takada-hidai01};
Israelian et al. \cite{israelian01}):
 (i) oxygen is reduced in the atmospheres of
giants because the convective envelope has mixed very oxygen-poor
material to the surface from the ON-cycled interior, and (ii)
the [O\,{\sc i}] line is partially filled in by emission.
These are suspicions without firm supporting observational evidence.
Reduction of oxygen by mixing must lead to a severe enhancement of the
nitrogen abundance, a result not suspected from examination of spectra
and not yet confirmed by quantitative analysis. If emission is the
culprit, is it not surprising that the [O\,{\sc i}] lines have never
been seen in emission in metal-poor giants, and that the
absorption line profile has not been reported as distorted or shifted
in radial velocity? 

A new way to approach the problem of the oxygen-to-iron ratio in
metal-poor stars is to determine the abundances from 
observations of the [O\,{\sc i}] and Fe\,{\sc ii} lines in
metal-poor ([Fe/H] $< -1$)
main sequence and subgiant stars. Mixing in these stars
is not predicted. Emission in the [O\,{\sc i}] lines would seem highly
improbable. The difficulty is that the [O\,{\sc i}] lines are
very weak in the spectra of such stars, and up to now only 4 dwarfs
(Spiesman \& Wallerstein \cite{spiesman91};
Spite \& Spite \cite{spite91}) and 2 subgiants
(Fulbright \& Kraft \cite{fulbright99};
Cayrel et al. \cite{cayrel01}; Israelian et al. \cite{israelian01})
have been
studied. It is our goal in this paper to provide a more
extensive study of the [O\,{\sc i}] 6300 \AA, the stronger of the 2 lines,
in metal-poor main sequence and subgiant
stars based on new VLT/UVES spectra. At the
same time, we determine the iron abundance from weak Fe\,{\sc ii}
lines, which makes the derived O/Fe ratio practically independent of
non-LTE effects, uncertainties in the assumed surface gravity, and only
weakly dependent on errors in the effective temperature, $T_{\rm eff}$.
Furthermore, we compare the \forOI\ based abundances with non-LTE
oxygen abundances derived from the \OI\ 7774\,\AA\ triplet.

\section{Observations and data reduction}
The oxygen and iron abundances have been derived from spectra obtained 
with UVES,
the Ultraviolet and Visual Echelle Spectrograph (Dekker et al. \cite{dekker00})
at the ESO VLT 8m
Kueyen telescope, during three observing runs:
April 8-12, 2000, July 22-24, 2000 and April 10-12, 2001.
The main purpose of the April observations was to determine beryllium
abundances from the \BeII\ lines at 3130\,\AA\ as observed in the blue arm
of UVES. % (Primas et al. \cite{primas02}).
Using a dichroic filter the spectral range 4800-6800\,\AA\ region was
observed simultaneously in the red arm of UVES with a resolution of 55\,000 and
about 4 pixels per spectral resolution element. The July 2000
observations were primarily conducted to determine the \lirat\
isotopic ratio from the profile of the \LiI\ resonance line at
6708~\AA. % (Asplund et al. \cite{asplund02a}).
The spectra were obtained with a narrow slit (0.3 arc sec) 
in combination with an image slicer and have a resolution of 120\,000 with
two pixels per spectral resolution element. Only the red arm of UVES was
used and the spectral ranges covered were 6000 - 7000~\AA\ with the EEV CCD and
7100-8200~\AA\ with the MIT chip.

The spectra have been reduced using MIDAS for the April spectra and IRAF
for the July spectra. Extensive testing has been performed between these
two data reduction packages, showing that the final spectra are indeed
of similar quality.
In both cases the reduction includes
definition of the echelle orders, background subtraction,
flat-field division, extraction of orders, wavelength calibration and
continuum normalization. The $S/N$ of the resulting spectra are very
high: 300-400 per pixel for the April spectra and 400-600 for the July
spectra. The continuum was fitted using cubic spline functions with
a wavelength scale of about 5\,\AA . As the spectrum of a metal-poor
star contains many good continuum regions this ensures that the continuum is 
well defined except in the vicinity of broad spectral features like the
H$\alpha$ line and the bandhead of telluric O$_2$ lines.  

Depending on the radial velocity of the star, the \forOI\ line may be blended
by telluric O$_2$ and H$_2$O lines. In order to be able to remove these
lines, several rapidly rotating
B-type stars were observed. The IRAF task {\em telluric}
was then used to 
optimize the fit between the telluric lines of the program star and the
B-type star by allowing a relative scaling in airmass and a
wavelength shift. The procedure worked well for the July 2000
observations due to the high resolution and a 
stable spectral PSF of UVES as defined by the output from the image slicer
and the 0.3 arcsec entrance slit, but was less succesfull for the April
observations, apparently because the PSF in this case is defined 
by a convolution of the entrance slit (0.7 arcsec) and the variable
seeing profile of the star. Hence, we were not able to use the \forOI\
line in the April spectra if it was significantly overlapped by
telluric lines.

\begin{figure}
\resizebox{\hsize}{!}{\includegraphics{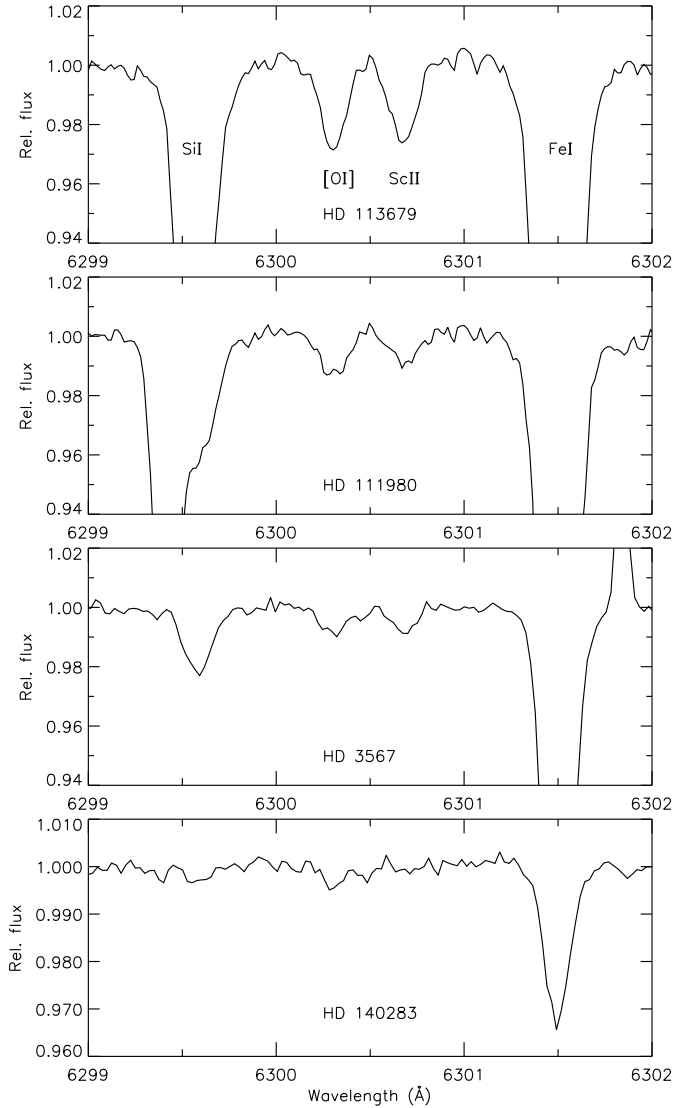}}
\caption{Representative spectra in the 6300~\AA\ region
arranged in order of decreasing strength of the \forOI\ line.
The spectra are corrected for Doppler shifts to the laboratory wavelengths
of the stellar lines. In the spectrum of \object{HD\,111980} a telluric O$_2$
line is blending the \SiI\ line and in the case of \object{HD\,3567}, the atmospheric
emission line of \forOI\ is seen to the right of the \FeI\ line.}
\label{fig:spectra}
\end{figure}

Figure \ref{fig:spectra} shows some representative
spectra. The first two are from the April 2000 observations and have 
$S/N$ close to 400, whereas the last two are from July 2000 and have $S/N 
\simeq 600$. We note that in the case of \object{HD\,140283} the \forOI\ line was
blended by a weak telluric H$_2$O line ($W \simeq 1$\,m\AA ). Care was taken
to remove this line by a proper scaling of the B-type calibration spectrum
based on numerous other H$_2$O lines, but the presence of the blending
H$_2$O line adds  to the uncertainty of the oxygen abundance
derived for \object{HD\,140283}.

\section{Stellar parameters}
On the basis of available Str\"{o}mgren photometry and other literature
data the programme stars were selected to lie rather close
to the turnoff region of halo stars and to span the
metallicity range $-3.0 < \feh < -0.6$. The oxygen abundance could not be
determined for all stars because either  the \forOI\ 6300 \AA\ line
was too weak to be detected or it was blended with telluric lines. Still,
we want to derive stellar parameters, \teff , \logg\ and \feh\ for
all program stars. These parameters will also be needed in connection with
the study of Li and Be abundances.

\subsection{Effective temperature}

\teff\ was determined from the $b-y$ and $V-K$ colour indices using the
IRFM calibrations of Alonso et al. (\cite{alonso96a}). Str\"omgren $uvby$-$\beta$
photometry
was taken from Schuster \& Nissen (\cite{schuster88}) for the large majority
of the stars supplemented with unpublished photometry of Schuster et al.
(\cite{schuster02}) for the remaining stars. The $(b-y)_0$ calibration of 
Schuster \& Nissen (\cite{schuster89}) (including a zero-point correction
of +0.005\,mag, Nissen \cite{nissen94}) was used to derive the interstellar
reddening excess. If $E(b-y) > 0.015$, the reddening is considered
significant and the V magnitude as well as the $m_1$ and the $c_1$ indices
are corrected
according to the relations $V_0$ = $V - 4.3 E(b-y)$, 
$m_0 = m_1 + 0.3 E(b-y)$ and $c_0 = c_1 - 0.2 E(b-y)$. These values and
the reddening estimates are given in Table 1.

\begin{table*}
\caption[ ]{Str\"{o}mgren photometry, colour excess, $V-K$ index,
\teff\ from $b-y$ and $V-K$, spectroscopic value of \feh , and
absolute magnitudes derived from the Str\"{o}mgren photometry and
from the Hipparcos parallax including an error corresponding to the
parallax error. If
$E(b-y) > 0.015$, the $V$ magnitudes and the photometric indices have been
corrected for interstellar absorption. For the two SB2 stars the photometric
value of \feh\ is given.}
\label{table1}
\setlength{\tabcolsep}{0.10cm}
\begin{tabular}{lrccccrcllccccc}
\noalign{\smallskip}
\hline
\noalign{\smallskip}
Star & $V_0$ & $(b-y)_0$ & $m_0$ & $c_0$ & $\beta$ & $E_{b-y}$ & $(V-K)_0$ & $T_{\rm eff}$ & $T_{\rm eff}$ &
 \feh & $M_{V,phot}$ & $M_{V,par}$ & $\pm \sigma$ & Note \\
%MA: not possible to fit in two columns with $T_{\rm eff}(b-y)$ and $T_{\rm eff}(V-K)$  
     & & & & & & & & $(b-y)$ & $(V-K)$ && & & & \\
\noalign{\smallskip}
\hline
\noalign{\smallskip}
      HD\,3567 &  9.255 & 0.332 & 0.087 & 0.334 & 2.598 & $-$0.002 &  1.369 &  6041\,K &  5958\,K & $-$1.16 &  3.86 &  4.16 & 0.31 &   \\
     HD\,19445 &  8.053 & 0.352 & 0.051 & 0.203 & 2.583 & $-$0.001 &  1.394 &  5877 &  5942 & $-$2.04 &  5.14 &  5.12 & 0.10 &   \\
     HD\,76932 &  5.801 & 0.354 & 0.117 & 0.297 & 2.581 & $-$0.017 &  1.445 &  5885 &  5829 & $-$0.86 &  4.42 &  4.16 & 0.04 &   \\
     HD\,97320 &  8.168 & 0.338 & 0.081 & 0.303 & 2.614 &  0.014 &  1.364 &  5985 &  5968 & $-$1.21 &  4.20 &  4.42 & 0.09 &   \\
     HD\,97916 &  9.209 & 0.293 & 0.104 & 0.407 & 2.638 &  0.002 &  1.173 &  6361 &  6296 & $-$0.86 &  3.59 &  3.64 & 0.35 &   \\
     HD\,99383 &  9.076 & 0.343 & 0.063 & 0.275 & 2.593 &  0.006 &  1.390 &  5945 &  5934 & $-$1.6: &  4.43 &  4.28 & 0.29 &   SB2 \\
    HD\,103723 &  9.947 & 0.329 & 0.104 & 0.306 & 2.625 &  0.027 &        &  6064 &       & $-$0.79 &  4.37 &  4.36 & 0.46 &   \\
    HD\,106038 & 10.179 & 0.342 & 0.092 & 0.264 & 2.583 & $-$0.018 &  1.408 &  5939 &  5897 & $-$1.37 &  4.69 &  4.99 & 0.36 &   \\
    HD\,111980 &  8.366 & 0.369 & 0.106 & 0.292 & 2.574 & $-$0.008 &  1.585 &  5777 &  5610 & $-$1.07 &  4.37 &  3.85 & 0.24 &   \\
    HD\,113679 &  9.627 & 0.386 & 0.136 & 0.320 & 2.579 &  0.017 &  1.560 &  5708 &  5649 & $-$0.66 &  4.21 &  3.80 & 0.42 &   \\
    HD\,116064 &  8.700 & 0.324 & 0.057 & 0.273 & 2.603 &  0.025 &  1.345 &  6100 &  6025 & $-$2.0: &  4.60 &  4.66 & 0.20 &   SB2 \\
    HD\,121004 &  9.031 & 0.395 & 0.140 & 0.268 & 2.588 &  0.012 &  1.557 &  5618 &  5653 & $-$0.75 &  4.89 &  5.15 & 0.18 &   \\
    HD\,122196 &  8.632 & 0.320 & 0.062 & 0.335 & 2.597 &  0.023 &        &  6137 &       & $-$1.70 &  3.75 &  3.58 & 0.29 &   \\
    HD\,126681 &  9.300 & 0.400 & 0.130 & 0.191 & 2.553 & $-$0.013 &  1.640 &  5532 &  5515 & $-$1.17 &  5.50 &  5.71 & 0.16 &   \\
    HD\,132475 &  8.377 & 0.360 & 0.075 & 0.277 & 2.578 &  0.041 &  1.463 &  5829 &  5808 & $-$1.45 &  4.39 &  3.55 & 0.23 &   \\
    HD\,140283 &  7.213 & 0.380 & 0.033 & 0.284 & 2.564 &  0.015 &  1.580 &  5734 &  5646 & $-$2.42 &  3.53 &  3.42 & 0.12 &   \\
    HD\,160617 &  8.733 & 0.347 & 0.051 & 0.331 & 2.584 &  0.011 &  1.408 &  5953 &  5909 & $-$1.79 &  3.27 &  3.42 & 0.31 &   \\
    HD\,166913 &  8.230 & 0.327 & 0.074 & 0.309 & 2.598 & $-$0.003 &        &  6069 &       & $-$1.56 &  4.18 &  4.26 & 0.14 &   \\
    HD\,175179 &  9.072 & 0.384 & 0.146 & 0.268 & 2.582 & $-$0.004 &  1.594 &  5690 &  5597 & $-$0.70 &  4.89 &  4.44 & 0.28 &   \\
    HD\,188510 &  8.834 & 0.416 & 0.100 & 0.163 & 2.553 &  0.010 &  1.669 &  5423 &  5481 & $-$1.61 &  5.83 &  5.85 & 0.10 &   \\
    HD\,189558 &  7.735 & 0.386 & 0.111 & 0.284 & 2.572 &  0.003 &  1.607 &  5665 &  5561 & $-$1.11 &  4.47 &  3.58 & 0.16 &   \\
    HD\,195633 &  8.472 & 0.343 & 0.121 & 0.368 & 2.606 &  0.018 &        &  6017 &       & $-$0.54 &  3.60 &  3.15 & 0.29 &   \\
    HD\,205650 &  9.052 & 0.375 & 0.093 & 0.231 & 2.569 & $-$0.005 &  1.490 &  5707 &  5759 & $-$1.17 &  5.02 &  5.40 & 0.14 &   \\
    HD\,213657 &  9.658 & 0.322 & 0.050 & 0.337 & 2.597 &  0.007 &        &  6144 &       & $-$1.94 &  3.55 &  3.43 & 0.59 &   \\
    HD\,298986 & 10.049 & 0.324 & 0.080 & 0.301 &       &        &  1.312 &  6081 &  6061 & $-$1.35 &  4.35 &  4.48 & 0.40 &   \\
    HD\,338529 &  9.370 & 0.308 & 0.045 & 0.366 & 2.600 &  0.006 &  1.260 &  6278 &  6200 & $-$2.34 &  3.22 &  3.57 & 0.46 &   \\
   CD$-30\degr18140$ &  9.859 & 0.302 & 0.053 & 0.340 & 2.606 &  0.021 &  1.214 &  6288 &  6260 & $-$1.91 &  3.93 &  4.18 & 0.46 &   \\
   CD$-35\degr14849$ & 10.568 & 0.321 & 0.040 & 0.293 & 2.603 &  0.010 &        &  6147 &       & $-$2.34 &  4.24 &       &      &   \\
   CD$-57\degr1633$ &  9.527 & 0.343 & 0.113 & 0.271 &       &        &  1.418 &  5944 &  5873 & $-$0.89 &  4.66 &  4.67 & 0.19 &   \\
    G\,013$-$009 &  9.916 & 0.292 & 0.054 & 0.369 & 2.609 &  0.019 &  1.219 &  6402 &  6275 & $-$2.30 &  3.68 &  3.71 & 0.59 &   \\
    G\,020$-$024 & 10.764 & 0.278 & 0.059 & 0.357 & 2.627 &  0.084 &  1.132 &  6469 &  6411 & $-$1.78 &  4.33 &       &      &   \\
    G\,075$-$031 & 10.521 & 0.333 & 0.093 & 0.330 & 2.597 & $-$0.005 &  1.421 &  6035 &  5870 & $-$1.04 &  3.96 &       &      &   \\
    G\,088$-$032 & 10.780 & 0.311 & 0.039 & 0.356 & 2.591 & $-$0.002 &  1.235 &  6251 &  6252 & $-$2.41 &  3.26 &       &      &   \\
    G\,126$-$062 &  9.478 & 0.330 & 0.063 & 0.327 & 2.588 & $-$0.004 &  1.455 &  6058 &  5824 & $-$1.58 &  3.75 &  4.11 & 0.37 &   \\
    G\,183$-$011 &  9.861 & 0.319 & 0.050 & 0.310 & 2.602 &  0.006 &  1.354 &  6170 &  6015 & $-$2.10 &  4.08 &       &      &   \\
    G\,271$-$162 & 10.264 & 0.306 & 0.055 & 0.355 & 2.602 &  0.020 &  1.275 &  6289 &  6170 & $-$2.31 &  3.61 &       &      &   \\
    LP\,815$-$43 & 10.774 & 0.272 & 0.054 & 0.376 & 2.623 &  0.032 &  1.140 &  6558 &  6471 & $-$2.72 &  4.17 &       &      &   \\
\hline
\end{tabular}
\end{table*}

The $K$ photometry was taken from Carney (\cite{carney83a}),
Alonso et al. (\cite{alonso94}) and The Two Micron All Sky Survey
(2MASS) data base
(Finlator et al. \cite{finlator00}). Furthermore, Dr. A. Alonso kindly
supplied unpublished
$K$ magnitudes for a few additional stars and Dr. G. Simon sent us 
$K$ photometry for a number of southern stars observed in connection
with the DENIS project (Epchtein et al. \cite{epchtein99}).
In general, the photometry of the various sources
agree within a typical scatter of $\pm 0.04$ mag for stars
in common
and no obvious systematic differences are seen. A straight mean value of
$K$ from the 
various sources has therefore been adopted. Table 1 lists the corresponding
$V-K$ index corrected for interstellar reddening
according to the relation $E(V-K) = 2.7 E(B-V) = 3.8 E(b-y)$ 
(Savage \& Mathis \cite{savage79}) if $E(b-y) > 0.015$.

When deriving \teff\ from $b-y$ it was noted that the calibration equation
of Alonso et al. (\cite{alonso96a}, Eq. 9) contains a quadratic term in
\feh , which introduces a tendency for \teff\ to diverge towards 
high values at very low metallicities. As pointed out by
Ryan et al. (\cite{ryan99}) this behaviour
(see their Fig. 5d) is likely to be a numerical artifact of the calibration.
We have investigated the problem
by comparing \teff\ derived from the Alonso et al. calibration with
original IRFM temperatures from Alonso et al. (\cite{alonso96b}) for a 
sample of 49 stars with Str\"{o}mgren photometry and $\feh < -1.5$.
Indeed, there is a trend that stars with $\feh < -2.0$ 
are assigned too high
 a \teff\  from the Alonso et al. calibration. To avoid this problem we have
simply adopted a lower limit of $\feh = -2.1$ in their Eq. (9).
The \teff\ calibration of $V-K$ (Alonso et al. \cite{alonso96a}, Eq. 7)
has also a square term in \feh . The
coefficient is, however, very small and the term is quite insignificant
even at the lowest metallicities, $\feh \simeq -3.0$.

\begin{figure}
\resizebox{\hsize}{!}{\includegraphics{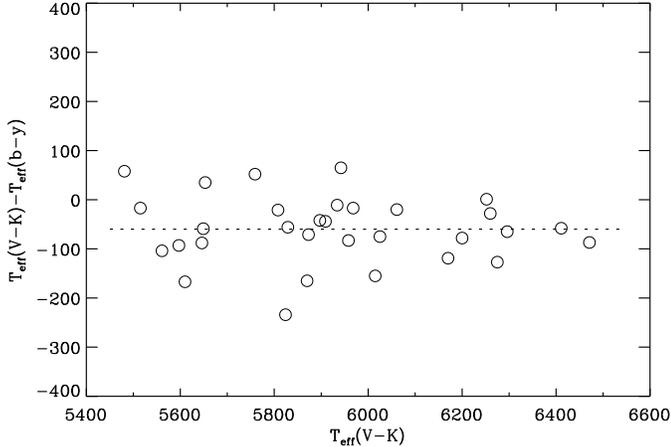}}
\caption{The difference between effective temperatures derived from $b-y$
and $V-K$ vs. $\teff (V-K)$.}
\label{fig:teff}
\end{figure}

The typical observational error of $b-y$ is 0.007 mag, which corresponds
to an error of $\pm 50$~K in \teff . The error of $V-K$ may be as high as
0.05 mag, but due to the high sensitivity of $V-K$ to \teff\ this also
leads to an error of $\pm 50$~K only.
An independent estimate of the error may be obtained
by comparing \teff\ derived from $b-y$ and $V-K$, respectively.
Table 1 lists the data and Fig. \ref{fig:teff} shows the comparison.
As seen, $\teff (b-y)$ is on the average 60~K higher than $\teff (V-K)$.
We have not found any explanation of this offset, but we note that
there are no significant correlations between the deviation
and \teff , absolute magnitude, or \feh . Hence, we adopt
the mean value of $\teff (b-y)$ and $\teff (V-K)$.
For the six stars missing $V-K$, a value $\teff (b-y) - 30$~K has
been adopted.

The rms scatter of the \teff\  difference shown in Fig. \ref{fig:teff}
is 68~K. This
corresponds to a statistical error of about $\pm 50$~K on 
$\teff (b-y)$ and $\teff (V-K)$ assuming that they have the same error.
If both $b-y$ and $V-K$ are
available we should then be able to
reach a statistical error of \teff\ of about $\pm 35$~K, which is comparable
to the accuracy claimed by Ryan et al. (\cite{ryan99}) in their study of
Li abundances of metal poor halo stars. This is, however, an optimistic
estimate of the error. Several of the stars, especially the most
metal-poor ones, are distant enough to be significantly reddened, and 
considering that the reddening estimate is hardly better than $E(b-y) \simeq
0.01$ an error of  70~K is introduced in both $\teff (b-y)$
and $\teff (V-K)$.
Furthermore, \teff\ may be affected by systematic errors,
which could depend on \feh\ and the position of the star in
the HR diagram. 

\subsection{Surface gravity}

The surface gravity was determined from the fundamental relation
\begin{eqnarray}
 \log \frac{g}{g_{\sun}} & = & \log \frac{\cal{M}}{\cal{M}_{\sun}}
 + 4 \log \frac{\teff}{T_{\rm eff,\sun}} + \\
    &   & 0.4 (M_{bol} - M_{bol,\sun})  \nonumber
\end{eqnarray}
where $\cal{M}$ is the mass of the star and $M_{bol}$ the absolute bolometric
magnitude.

\begin{figure}
\resizebox{\hsize}{!}{\includegraphics{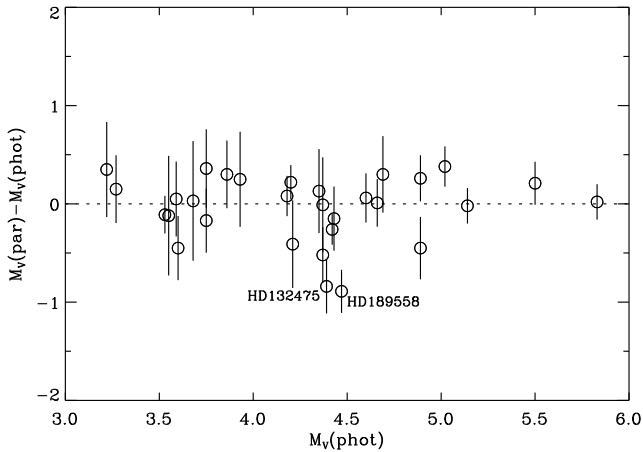}}
\caption{The difference between the absolute visual magnitude determined
from Str\"{o}mgren photometry and from the Hipparcos parallax.
The error bars correspond to the parallax error and
an estimated error of $M_{V,phot}$ of 0.15 mag.}
\label{fig:Mv}
\end{figure}

The absolute visual magnitude \Mv\ was determined from a new calibration
of the Str\"{o}mgren indices derived by Schuster et al.
(\cite{schuster02}) on the basis of Hipparcos parallaxes, and 
also directly from the Hipparcos parallax (ESA \cite{esa97})
if available with
an accuracy $\sigma (\pi) / \pi < 0.3$. Column 12 and 13 of
Table 1 lists the photometric and the parallax based values of \Mv\
and in Fig. \ref{fig:Mv} the values are compared. As seen
the agreement is quite satisfactory considering the estimated error bars.
For two stars, \object{HD\,132475} and \object{HD\,189558}, $M_{V,par}$ has a 3$\sigma$
deviation from $M_{V,phot}$ of the order $-0.8$ mag, which suggests that these
stars are binaries consisting of two almost equal components. There is, however,
no indication of double or asymmetric lines in their spectra,
so we reject the binary hypothesis and consider the deviation in 
Fig. \ref{fig:Mv} as accidental. Two other stars, \object{HD\,99383}
and \object{HD\,116064}, have
double lines, clearly indicating the presence of a fainter and cooler component
to the main star. These spectroscopic binaries (SB2) 
stars were excluded from further analysis.

In calculating \logg\ we adopted the mean value of $M_{V,phot}$ and $M_{V,par}$
if the parallax is available, otherwise $M_{V,phot}$.
The bolometric correction is taken from Alonso et al. (\cite{alonso95}, 
Table 4). The mass was derived by interpolating in the \Mv -- \logteff\
diagram between the $\alpha$-element enhanced evolutionary tracks of 
VandenBerg et al. (\cite{vandenberg00}).
The internal precision of the mass is estimated to
be better than $\pm 0.05 \,\, \cal{M}_{\sun}$.

The error of the derived value of \logg\ is dominated by the error
of \Mv ; $\sigma (\Mv) = \pm 0.20$ corresponds to
$\sigma (\logg) = \pm 0.08$~dex. In addition there could be a small
systematic error
of \logg\ but in view of the good overall agreement between 
$M_{V,phot}$ and $M_{V,par}$ this systematic error is probably less
than $\pm 0.10$~dex. Altogether we estimate that the error of \logg\
is on the order of $\pm 0.15$ dex.

The final values of \teff\ and \logg\ are given in Table 2 together
with values for \feh\ and the microturbulence as derived from \FeII\ lines
(Sect. 4). In addition, the absolute magnitude and the mass are given.
As the calibration of \teff\ and hence also \logg\ depend somewhat on \feh\
the determination of the three parameters is an iterative process.

\begin{table}
\caption[ ]{ The derived values of effective temperature,
surface gravity, metal abundance,
microturbulence, absolute magnitude and mass.}
\label{table2}
\setlength{\tabcolsep}{0.10cm}
\begin{tabular}{lcccccc}
\noalign{\smallskip}
\hline
\noalign{\smallskip}
Star & $\teff$ & $\logg$ & \feh & $\xi_{\rm micro}$ & $M_V$ & Mass  \\
     & [K]     & [cgs]   &      & [km/s] &      &[$\cal{M}_{\odot}$]   \\
\noalign{\smallskip}
\hline
\noalign{\smallskip}
   \object{HD\,3567} &    6000 &   4.07 &  $-$1.16 &   1.5 &    4.01 &   0.82  \\ 
  \object{HD\,19445} &    5910 &   4.40 &  $-$2.04 &   1.3 &    5.13 &   0.70  \\ 
  \object{HD\,76932} &    5857 &   4.15 &  $-$0.86 &   1.2 &    4.29 &   0.85  \\ 
  \object{HD\,97320} &    5976 &   4.16 &  $-$1.21 &   1.3 &    4.31 &   0.78 \\ 
  \object{HD\,97916} &    6328 &   4.11 &  $-$0.86 &   1.5 &    3.61 &   1.02 \\ 
 \object{HD\,103723} &    6034 &   4.26 &  $-$0.79 &   1.3 &    4.36 &   0.88 \\ 
 \object{HD\,106038} &    5918 &   4.31 &  $-$1.37 &   1.2 &    4.84 &   0.70 \\ 
 \object{HD\,111980} &    5694 &   3.99 &  $-$1.07 &   1.2 &    4.11 &   0.79 \\ 
 \object{HD\,113679} &    5678 &   4.04 &  $-$0.66 &   1.2 &    4.00 &   0.96 \\ 
 \object{HD\,121004} &    5635 &   4.34 &  $-$0.75 &   1.1 &    5.02 &   0.80 \\ 
 \object{HD\,122196} &    6107 &   3.93 &  $-$1.70 &   1.5 &    3.66 &   0.78 \\ 
 \object{HD\,126681} &    5524 &   4.48 &  $-$1.17 &   0.7 &    5.61 &   0.70 \\ 
 \object{HD\,132475} &    5818 &   3.95 &  $-$1.45 &   1.4 &    3.97 &   0.75 \\ 
 \object{HD\,140283} &    5690 &   3.69 &  $-$2.42 &   1.5 &    3.47 &   0.75 \\ 
 \object{HD\,160617} &    5931 &   3.77 &  $-$1.79 &   1.5 &    3.35 &   0.82 \\ 
 \object{HD\,166913} &    6039 &   4.11 &  $-$1.56 &   1.3 &    4.22 &   0.73  \\ 
 \object{HD\,175179} &    5643 &   4.20 &  $-$0.70 &   1.0 &    4.66 &   0.80  \\ 
 \object{HD\,188510} &    5452 &   4.53 &  $-$1.61 &   0.8 &    5.84 &   0.68  \\ 
 \object{HD\,189558} &    5613 &   3.91 &  $-$1.11 &   1.2 &    4.03 &   0.76  \\ 
 \object{HD\,195633} &    5987 &   3.95 &  $-$0.54 &   1.4 &    3.38 &   1.10  \\ 
 \object{HD\,205650} &    5733 &   4.39 &  $-$1.17 &   1.0 &    5.21 &   0.70  \\ 
 \object{HD\,213657} &    6114 &   3.85 &  $-$1.94 &   1.5 &    3.49 &   0.77  \\ 
 \object{HD\,298986} &    6071 &   4.21 &  $-$1.35 &   1.3 &    4.41 &   0.76 \\ 
 \object{HD\,338529} &    6239 &   3.87 &  $-$2.34 &   1.5 &    3.40 &   0.79  \\ 
\object{CD\,$-30\degr18140$} &  6274 &   4.12 &  $-$1.91 &   1.5 &    4.06 &   0.75  \\ 
\object{CD\,$-35\degr14849$} &  6117 &   4.12 &  $-$2.34 &   1.5 &    4.24 &   0.70  \\ 
\object{CD\,$-57\degr1633$}  &  5909 &   4.32 &  $-$0.89 &   1.2 &    4.66 &   0.84  \\ 
 \object{G\,013-009} &    6338 &   4.00 &  $-$2.30 &   1.5 &    3.70 &   0.76  \\ 
 \object{G\,020-024} &    6440 &   4.30 &  $-$1.78 &   1.5 &    4.33 &   0.78  \\ 
 \object{G\,075-031} &    5952 &   4.07 &  $-$1.04 &   1.3 &    3.96 &   0.87  \\ 
 \object{G\,088-032} &    6251 &   3.81 &  $-$2.41 &   1.5 &    3.26 &   0.80  \\ 
 \object{G\,126-062} &    5941 &   3.98 &  $-$1.58 &   1.3 &    3.93 &   0.76  \\ 
 \object{G\,183-011} &    6092 &   4.04 &  $-$2.10 &   1.5 &    4.08 &   0.70  \\ 
 \object{G\,271-162} &    6230 &   3.93 &  $-$2.31 &   1.5 &    3.61 &   0.75  \\ 
 \object{LP\,815-43} &    6514 &   4.23 &  $-$2.72 &   1.5 &    4.17 &   0.75  \\ 
\hline
\end{tabular}
\end{table}

\section{O and Fe abundances}

The determination of abundances is based
on $\alpha$-element enhanced
(\alphafe = +0.4, $\alpha$ = C, O, Ne, Mg, Si, S, Ca, and Ti) 
1D model atmospheres with \teff , \logg\ and \feh\ values
as given in Table 2 and a microturbulence of 1.0 \kmprs .
The models were computed with the MARCS code using updated continuous
opacities (Asplund et al. \cite{asplund97}) and including UV line blanketing
by millions of absorption lines.
LTE is assumed both in constructing the models and in deriving abundances.
We first discuss the determination of Fe abundances from \FeII\ lines
and then O abundances derived from the \forOI\  6300 \AA\ line and the
\OI\ triplet at 7774 \AA .

\subsection{Iron abundance
\label{s:FeII}}
By inspecting the spectra of typical program stars as well as the
solar spectrum, 13 apparently unblended \FeII\ lines from
Bi\'{e}mont et al. (\cite{biemont91},
Table 2)  were selected as suitable for determining the Fe abundance.
The lines are listed in Table 3. The log$gf$ values given are those of
Bi\'{e}mont et al., i.e. based on theoretical transition probabilities and
a normalization ($-0.05$ dex) from lifetime measurements. 
The lifetime measurements reported by Bi\'{e}mont et al. agree well with
the re-analysis by Allende Prieto et al. (\cite{prieto02})
of published experimental data, including more recent results.

\begin{table*}
\caption[ ]{ List of \FeII\ lines used for determining the iron
abundances of the stars. Equivalent widths as observed in the solar
flux spectrum and in the spectra of \object{HD\,189558},
\object{HD\,160617} and \object{HD\,140283}
are given together with the derived abundances for each line.}
\label{table3}
\setlength{\tabcolsep}{0.10cm}
\begin{tabular}{cccrrrr}
\noalign{\smallskip}
\hline
\noalign{\smallskip}
 Wavelength & exc.pot. & log$gf$ & Solar flux &  HD\,189558  &  HD\,160617 & HD\,140283 \\
     A      &    eV    &         &\hspace{0.4cm} $W$(m\AA) log$\epsilon$(Fe) & 
  \hspace{0.5cm}  $W$(m\AA) \hspace{0.3cm} \feh &
  \hspace{0.5cm}  $W$(m\AA) \hspace{0.3cm} \feh &
 \hspace{0.5cm} $W$(m\AA) \hspace{0.3cm} \feh \\
\noalign{\smallskip}
\hline
\noalign{\smallskip}
   4993.35  &   2.81 & $-$3.54 &  40.5 \hspace{0.3cm}  7.33 &  13.5 \hspace{0.3cm} $-$1.10 &  3.8 \hspace{0.3cm} $-$1.81 & \\   
   5100.66  &   2.81 & $-$4.19 &  19.6 \hspace{0.3cm}  7.46 &   4.5 \hspace{0.3cm} $-$1.12 &  1.2 \hspace{0.3cm} $-$1.81 & \\
   5197.58  &   3.23 & $-$2.28 &  88.4 \hspace{0.3cm}  7.52 &  53.3 \hspace{0.3cm} $-$1.13 & 29.7 \hspace{0.3cm} $-$1.81
& 10.2 \hspace{0.3cm} $-$2.47 \\
   5234.62  &   3.22 & $-$2.20 &  90.5 \hspace{0.3cm}  7.46 &  56.4 \hspace{0.3cm} $-$1.09 & 33.4 \hspace{0.3cm} $-$1.76
& 12.7 \hspace{0.3cm} $-$2.39 \\
   5325.55  &   3.22 & $-$3.27 &  43.0 \hspace{0.3cm}  7.50 &  15.1 \hspace{0.3cm} $-$1.09 &  4.9 \hspace{0.3cm} $-$1.76 & \\
   5414.08  &   3.22 & $-$3.80 &  27.0 \hspace{0.3cm}  7.65 &   7.2 \hspace{0.3cm} $-$1.09 &  2.5 \hspace{0.3cm} $-$1.68 & \\
   5425.26  &   3.20 & $-$3.42 &  44.0 \hspace{0.3cm}  7.66 &  14.6 \hspace{0.3cm} $-$1.14 &  4.4 \hspace{0.3cm} $-$1.84
&  1.4 \hspace{0.3cm} $-$2.42 \\
   6084.10  &   3.20 & $-$3.86 &  21.3 \hspace{0.3cm}  7.52 &   5.5 \hspace{0.3cm} $-$1.07 &  1.4 \hspace{0.3cm} $-$1.80 & \\
   6149.25  &   3.89 & $-$2.77 &  38.0 \hspace{0.3cm}  7.51 &  11.8 \hspace{0.3cm} $-$1.10 &  3.9 \hspace{0.3cm} $-$1.76
&  1.2 \hspace{0.3cm} $-$2.33 \\
   6247.56  &   3.89 & $-$2.38 &  56.2 \hspace{0.3cm}  7.53 &  23.4 \hspace{0.3cm} $-$1.12 &  8.7 \hspace{0.3cm} $-$1.80 
&  2.1 \hspace{0.3cm} $-$2.49 \\
   6416.93  &   3.89 & $-$2.79 &  42.4 \hspace{0.3cm}  7.62 &  11.8 \hspace{0.3cm} $-$1.19 &  4.2 \hspace{0.3cm} $-$1.83 & \\
   6432.68  &   2.89 & $-$3.76 &  43.0 \hspace{0.3cm}  7.65 &  14.8 \hspace{0.3cm} $-$1.13 &  4.6 \hspace{0.3cm} $-$1.81
&  1.3 \hspace{0.3cm} $-$2.46 \\
   6456.39  &   3.90 & $-$2.13 &  66.5 \hspace{0.3cm}  7.50 &  33.0 \hspace{0.3cm} $-$1.09 & 14.2 \hspace{0.3cm} $-$1.77
&  4.3 \hspace{0.3cm} $-$2.39 \\
\noalign{\smallskip}
\hline
\noalign{\smallskip}
  Average:  &        &       &         7.53  &     $-$1.11    &  $-$1.79 & $-2.42$  \\
  Dispersion:&       &       &         0.08  &        0.03    &     0.04 &   0.06 \\
\hline
\end{tabular}
\end{table*}

Using the MARCS
solar model and assuming a microturbulence of 1.15~\kmprs\
the equivalent widths of the lines as measured in
the solar flux spectrum of Kurucz et al. (\cite{kurucz84}) were first
analyzed. The results
are given in Table 3. The average logarithmic solar abundance, 
log$\epsilon$(Fe) = 7.53,
is close to the commonly adopted meteoritic value of 7.50
(Grevesse \& Sauval \cite{grevesse98}), but the dispersion is
quite high (0.08 dex). We note that there is no significant
correlation between the Fe abundance and the equivalent width of the
lines suggesting that the chosen value of the microturbulence is
approximately correct. Furthermore, it is seen that when determining
differential Fe abundances for each line for three representative stars,
\object{HD\,140283}, \object{HD\,160617} and \object{HD\,189558}, an amazingly 
small line-to-line scatter of \feh\ (0.03 - 0.06 dex) is obtained.
This suggests that the scatter 
of the solar log$\epsilon$(Fe) values is mainly
due to errors in the $gf$-values. By 
working differentially with respect to the Sun this problem is avoided.
For the whole set of program stars the scatter in [Fe/H] ranges from
0.03 to about 0.08 dex, and since at least four \FeII\ lines are measured even 
in the most metal-poor stars (see Table A1
\footnote{Table A1 with equivalent widths of all lines measured
is only available in electronic form at the CDS
via anonymous ftp to cdsarc.u-strasbg.fr (130.79.128.5) or
via http://cdsweb.u-strasbg.fr/Abstract.html}),
the statistical error of \feh\ is less than 0.05 dex. 

In calculating abundances from the \FeII\ lines we adopted the
Uns\"{o}ld (\cite{unsold55}) approximation to the Van der Waals
interaction constant with an enhancement factor $E_{\gamma} = 2.0$.
We note that a decrease of $E_{\gamma}$ to 1.0 leads to an increase of the
solar iron abundance of +0.04 dex, whereas the \FeII\ lines in the stars
are weak enough to make the derived iron abundance practically independent
of $E_{\gamma}$. Hence, the adopted value of $E_{\gamma}$ has
no more than a small
systematic effect on the metallicity scale of the metal-poor stars
and the derived [O/Fe] values.

In the more metal-poor stars, $\feh < -1.5$, the \FeII\ lines are 
weak and hence the derived metallicity is practically independent of
the microturbulence. For such stars we have assumed 
$\xi_{\rm micro} = 1.5$ \kmprs . For the more metal-rich stars
$\xi_{\rm micro}$ has been derived by requesting that the derived \feh\ values
should be independent of equivalent width. The two medium strong lines,
5197.58 \AA\ and  5234.62 \AA,  get high weight in this
determination. The error of $\xi_{\rm micro}$ is about $\pm 0.2$~\kmprs . Due to
the dominance of weaker lines the corresponding error of the mean value
of \feh\ is negligible. It should be noted, by the way, that the
derived microturbulence increases in a systematic way along the
main sequence of stars in the HR-diagram, i.e. from about 0.8~\kmprs\
at $\teff \simeq 5500$~K to $\sim 1.5$~\kmprs\
for turnoff stars at $\teff \simeq 6300$~K.

The error of \feh\ induced by the uncertainty in \teff\  is small,
i.e. less than 0.02 dex. An error in
\logg\ ($\pm 0.15$) corresponds, however, to an error of $\pm 0.06$~dex in
\feh . Added (quadratically) to the error indicated by the line-to-line
scatter from equivalent width measurements and the $gf$-values,
we get  a typical
error of $\pm 0.08$ dex for \feh .

\subsection{Oxygen abundance from the \forOI\ 6300.3 \AA\ line
\label{s:[OI]}}

The equivalent width of the 6300.3 \AA\ spectral feature  
was measured relative to the continuum
defined by the cubic spline fitting procedure described in Sect. 2 
and by integrating the line depth 
over an interval of 0.35\AA\ centered on 6300.31~\AA . 
From the photon statistics we estimate an error
$\sigma (W) \simeq \pm 0.3$~m\AA\ for the April spectra and
$\sigma (W) \simeq \pm 0.2$~m\AA\ for the July spectra.
As described later this is confirmed by a $\chi ^2$
fitting between observed and synthetic spectra.
The total equivalent width of the  6300.3\,\AA\ feature is given
in Table 4 for all stars in which the line is detectable ($W \geq 0.4$\,m\AA ).

As recently shown by Allende Prieto et al. (\cite{prieto01}), the
\forOI\ line at 6300.304~\AA\ is significantly blended 
by a \NiI\ line at 6300.339~\AA\ with $\chi = 4.27$~eV.
From a  3D model atmosphere synthesis of the  6300.3 \AA\   feature
in the solar flux spectrum of Kurucz et al. (\cite{kurucz84}) 
they estimate the oscillator strength of the Ni line to be log\,$gf = -2.31$
for an adopted solar Ni abundance of log$\epsilon$(Ni) = 6.25.
The corresponding (3D) oxygen abundance of the Sun is log$\epsilon$(O) = 8.69
when using the accurate oscillator strength of the \forOI\ line,
log$gf = -9.72$ (Allende Prieto et al. \cite{prieto01}). 
%(Grevesse \& Sauval 1998) is in excellent agreement with a similar 3D analysis
%of the OH IR vibration-rotation and pure rotation lines (Asplund et al. 2002b), 
%strongly supporting the existence of the Ni blend in the \forOI\ line.

The  6300.3 \AA\   feature in the solar flux spectrum has an 
equivalent width of 5.5~m\AA . Subtracting the calculated equivalent width of
the Ni line (1.4~m\AA ), we determine a (1D) oxygen abundance of the
Sun, log$\epsilon$(O) = 8.74, in quite good agreement with the 3D
result of Allende Prieto et al.,
but  substantially lower than previous determinations that did not
take the Ni line into account. 
Indeed, when accounting for the difference of $-$0.04\,dex between 3D
and {\sc marcs} model atmosphere for the solar O abundance, our analysis
agrees almost perfectly with that of Allende Prieto et al. (\cite{prieto01}).

Adopting now log$gf = -2.31$ for the \NiI\ line and assuming that the
Ni abundance scales proportional to the Fe abundance 
(Edvardsson et al. \cite{edvardsson93}),
the equivalent width of the Ni line (see Table 4) has been calculated
and subtracted from
the total equivalent width of the  6300.3 \AA\ feature. The resulting
$W(\forOI)$ is used to determine the
oxygen abundances of the stars relative to that of the Sun, i.e. [O/H]. 
As seen from the table, the Ni line is rather weak in the metal poor
stars and its inclusion decreases the derived oxygen abundance
by less than 0.04 dex for stars with $\feh < -1.0$. For the Sun, however,
the effect of including the \NiI\ line is a decrease of the oxygen
abundance by 0.13~dex. Hence, the net effect of the Ni line
for metal-poor halo stars is an increase of [O/Fe] with about 0.1~dex.

\begin{table}
\caption[ ]{Equivalent widths, $W_{\rm tot}$, of the 6300.3 \AA\ absorption
feature together with calculated values for the blending
\NiI\ line, $W_{\rm Ni}$,
and values of log$\epsilon$(O) and [O/Fe] as derived from 1D
model atmospheres. The last two columns
give the $\pm 1\sigma$ errors of [O/Fe].} 
\label{table4}
\setlength{\tabcolsep}{0.10cm}
\begin{tabular}{rrcccccc}
\noalign{\smallskip}
\hline
\noalign{\smallskip}
 Star       & \feh     & $W_{\rm tot}$  & $W_{\rm Ni}$  & log$\epsilon$(O) & [O/Fe] & $+\sigma$ & $-\sigma$ \\
            &          &   m\AA       &  m\AA       &       &        &           &           \\
\noalign{\smallskip}
\hline
\noalign{\smallskip}
Sun (flux)  &  0.00    & 5.5  & 1.4  &  8.74 &  0.00  &        &       \\
HD\,3567    & $-$1.16  & 1.7  & 0.1  &  8.00 &  0.42  & +0.08  & $-$0.10 \\
HD\,76932   & $-$0.86  & 3.3  & 0.2  &  8.32 &  0.44  & +0.06  & $-$0.06 \\
HD\,97320   & $-$1.21  & 1.2  & 0.1  &  7.86 &  0.33  & +0.11  & $-$0.15 \\
HD\,103723  & $-$0.79  & 1.7  & 0.2  &  8.13 &  0.18  & +0.09  & $-$0.10 \\
HD\,111980  & $-$1.07  & 2.1  & 0.1  &  7.96 &  0.29  & +0.08  & $-$0.09 \\  
HD\,113679  & $-$0.66  & 4.5  & 0.4  &  8.39 &  0.31  & +0.05  & $-$0.05 \\
HD\,121004  & $-$0.75  & 3.1  & 0.3  &  8.30 &  0.31  & +0.06  & $-$0.07 \\
HD\,126681  & $-$1.17  & 1.8  & 0.1  &  8.03 &  0.46  & +0.09  & $-$0.10 \\
HD\,132475  & $-$1.45  & 1.5  & 0.1  &  7.78 &  0.49  & +0.09  & $-$0.11 \\
HD\,140283  & $-$2.42  & 0.5  & 0.0  &  7.09 &  0.77  & +0.15  & $-$0.22 \\
HD\,160617  & $-$1.79  & 0.6  & 0.0  &  7.37 &  0.42  & +0.19  & $-$0.30 \\
HD\,166913  & $-$1.56  & 0.6  & 0.0  &  7.57 &  0.39  & +0.18  & $-$0.30 \\
HD\,189558  & $-$1.11  & 3.0  & 0.1  &  7.06 &  0.43  & +0.06  & $-$0.07 \\
HD\,195633  & $-$0.54  & 3.0  & 0.4  &  8.28 &  0.08  & +0.07  & $-$0.07 \\
HD\,205650  & $-$1.17  & 1.6  & 0.1  &  8.00 &  0.43  & +0.09  & $-$0.11 \\
HD\,213657  & $-$1.94  & 0.4  & 0.0  &  7.31 &  0.51  & +0.18  & $-$0.30 \\
HD\,298986  & $-$1.35  & 0.8  & 0.0  &  7.77 &  0.38  & +0.15  & $-$0.21 \\
G\,075$-$031  & $-$1.04  & 1.2  & 0.1 & 7.85 &  0.15  & +0.11  & $-$0.15 \\
\hline
\end{tabular}
\end{table}

The effect of uncertainties in \teff\ and \logg\ on \ofe\ is
small; $\sigma (\teff) = \pm 70$~K corresponds to
$\sigma (\ofe ) = \pm 0.035$~dex, and most important, the effect
of varying the gravity is very nearly the same on log$\epsilon$(O)
and log$\epsilon$(Fe), when
the two abundances are derived from the \forOI\ line and \FeII\ lines,
respectively. For example, an increase of the gravity of \object{HD\,132475} by
as much as 0.3 dex changes [O/Fe] by less than 0.01 dex.
Hence, [O/Fe] is not affected by errors in \logg .
In practice,  the error budget of \ofe\ is totally dominated by
the error of the equivalent width of the \forOI\ line. In order to
check our estimate of this error source we have made a $\chi^2$ analysis of
a comparison of synthetic spectra with some representative observed spectra.

\begin{figure}
\resizebox{\hsize}{!}{\includegraphics{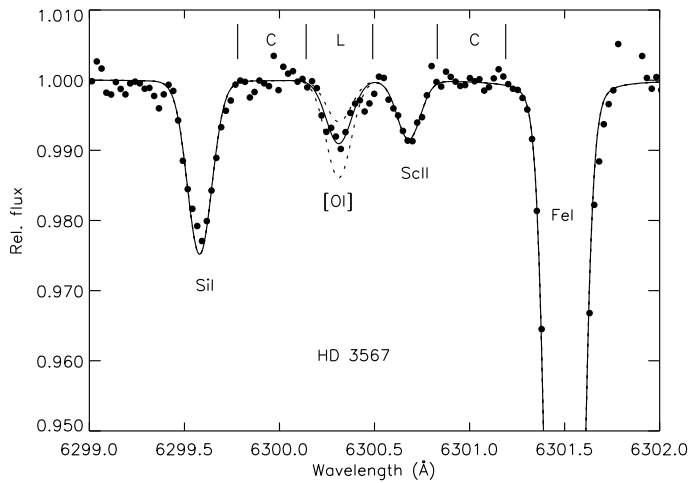}}
\caption{1D model atmosphere synthesis of the observed spectrum of
\object{HD\,3567} ($S/N \simeq 540$). The $\chi^2_{min}$ is obtained for
log$\epsilon$(O) = 7.95 (full drawn line). The two dashed lines correspond
to oxygen abundances changed with $\pm 0.2$~dex. The line and continuum 
regions used for the $\chi^2$ analysis are indicated. Note, that the two
deviating points to the right of the \FeI\ line are due to telluric \forOI\
emission.}
\label{fig:fit}
\end{figure}

The synthetic spectra are computed for the spectral region 6299 -- 6302~\AA .
In addition to the \forOI\ and \NiI\ lines, \SiI\
6299.58 \AA, \ScII\ 6300.68 \AA, and \FeI\ 6301.50 \AA\ 
are included, see Fig. \ref{fig:fit}.
The log$gf$ values of these lines were set to get agreement between their
calculated and observed equivalent widths, but they have practically no
influence on the $\chi^2$ analysis, because the spectral regions of these
lines are not included. The synthetic spectra are
folded with a Gaussian profile representing the instrumental, macroturbulent
and rotational broadening. The FWHM of this Gaussian is determined 
from the \FeI\  6301.50 \AA\   line, e.g. 5.8~\kmprs\ in the case
of \object{HD\,3567}. The $\chi^2$ analysis is
then performed over the spectral regions indicated on Fig. 
4, i.e. a total of 42 datapoints, of which the $C$ regions control the
continuum setting and the $L$ region determines the O abundance.
For each chosen oxygen abundance the
continuum setting is optimized to get the smallest possible $\chi^2$.
In the case of \object{HD\,3567} the minimum $\chi^2$ is obtained for 
log$\epsilon$(O) = 7.95 $\pm 0.05$. The reduced $\chi^2$ is 0.88, and  
the quoted (one-sigma) error corresponds to $\Delta \chi^2 = 1$.
This oxygen abundance is 0.05 dex lower than the value
determined from the measured equivalent width of the  6300.3 \AA\
feature, but well within the estimated error bars. A similar $\chi^2$ analysis
was performed for \object{HD\,160617} (July 2000 spectrum)
and \object{HD\,132475} and \object{HD\,189558}
(April 2000) confirming in all cases the oxygen abundances found
from the measured equivalent widths of the  6300.3  \AA\ spectral feature.

\subsection{Oxygen abundance from the \OI\ triplet
\label{s:triplet}}

\begin{figure}
\resizebox{\hsize}{!}{\includegraphics{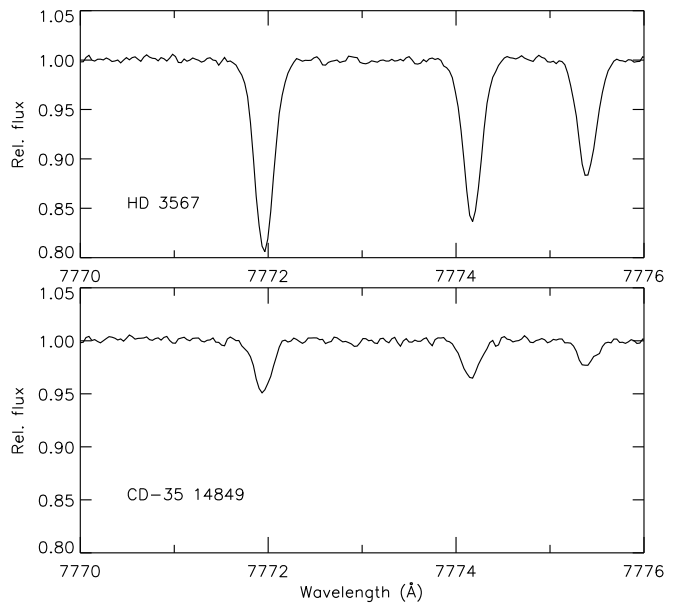}}
\caption{Spectra of two stars in the \OI\ triplet region}
\label{fig:triplet}
\end{figure}

For the 15 stars belonging to the July 2000 program the \OI\ triplet at 
7774\,\AA\ was included in the UVES spectra. The triplet occurs in
two consecutive echelle orders each with a S/N of 200-300 in the
wavelength region of the triplet. Hence, two sets of equivalent widths
could be measured and from the comparison, the error of the mean equivalent
widths (given in Table 5) is estimated to be less than 1\,m\AA .
The high quality of the spectra is evident from Fig. \ref{fig:triplet},
which shows the
average spectra of the triplet for two representative stars.

From the  equivalent widths of the triplet,
LTE abundances were first derived  using the model atmospheres employed in the
\forOI\ analysis. The adopted $gf$-values are  those
recommended by Wiese et al. (\cite{wiese96}): $\log gf$  = 0.369, 0.223, and
0.002. For a given star the three \OI\ lines provide the same
oxygen abundance to within  about 0.03 dex except in the case of \object{LP\,815-43},
for which the dispersion is around 0.09 dex due to the faintness of the
lines. Estimated uncertainties of $\pm$ 70 K in $T_{\rm eff}$, and
$\pm$ 0.15 in $\log g$ correspond to $\Delta\log\epsilon$(O) of
$\pm$ 0.05 and  $\pm$ 0.06, respectively.
Uncertainties arising from  the microturbulence
and [Fe/H] may be neglected. We estimate the total error of 
$\log\epsilon$(O) to be about $\pm$ 0.1 dex, whereas the error of \ofe\ is in fact
a bit smaller, because the gravity induced errors on the O and Fe abundances
nearly cancel.

\begin{table}
\caption[ ]{Equivalent widths of the 7771.2, 7774.2, and
7775.4\,\AA\ lines in the \OI\ triplet
and the derived mean LTE and non-LTE oxygen abundances. The last column
gives [O/Fe] based on the non-LTE oxygen abundance.}
\label{table5}
\setlength{\tabcolsep}{0.10cm}
\begin{tabular}{rcrrrccc}
\noalign{\smallskip}
\hline
\noalign{\smallskip}
 Star       & \feh     & $W_1$  & $W_2$  & $W_3$  & 
log$\epsilon$(O) & log$\epsilon$(O) & [O/Fe]  \\
            &          &   m\AA       &  m\AA     &  m\AA   & 
{\tiny LTE} & {\tiny n-LTE} & {\tiny n-LTE}   \\
\noalign{\smallskip}
\hline
  HD\,3567 & $-$1.16 & 49.2&41.0&29.5 & 8.14& 7.94& 0.36 \\
   HD\,19445 & $-$2.04 & 13.9& 9.7& 6.3 & 7.49& 7.39& 0.69 \\
  HD\,106038 & $-$1.37 & 39.5&32.1&22.9 & 8.10& 7.94& 0.57 \\
  HD\,140283 & $-$2.42 &  7.1& 5.8& 3.2 & 7.11& 7.02& 0.70 \\
  HD\,160617 & $-$1.79 & 19.9&14.9& 9.7 & 7.46& 7.34& 0.39 \\
  HD\,213657 & $-$1.94 & 26.5&20.7&14.0 & 7.56& 7.41& 0.61 \\
  HD\,338529 & $-$2.34 & 16.7&13.3& 7.4 & 7.20& 7.06& 0.66 \\
  $-30\degr18140$ & $-$1.91 & 27.2&21.5&14.7 & 7.56& 7.41& 0.58 \\
  $-35\degr14849$ & $-$2.34 & 11.1& 8.0& 5.2 & 7.15& 7.04& 0.64 \\
  G\,013$-$009 & $-$2.30 & 14.8&12.4& 7.8 & 7.15& 7.01& 0.57 \\
  G\,020$-$024 & $-$1.78 & 26.8&20.6&13.6 & 7.49& 7.34& 0.38 \\
  G\,075$-$031 & $-$1.04 & 55.7&44.4&34.3 & 8.28& 8.06& 0.36 \\
  G\,126$-$062 & $-$1.58 & 38.4&30.1&21.0 & 7.96& 7.79& 0.63 \\
  G\,271$-$162 & $-$2.31 & 13.2& 9.5& 5.9 & 7.08& 6.96& 0.53 \\
  LP\,815$-$43 & $-$2.72 &  6.0& 4.7& 3.9 & 6.71& 6.46& 0.43 \\
\hline
\end{tabular}
\end{table}
~

It is well established that the line formation process of the \OI\ triplet in late-type
stars is affected by departures from LTE (e.g. Kiselman 2001). To account for these
important non-LTE effects, we have performed statistical equilibrium calculations
using version 2.2 of the publicly available code MULTI (Carlsson 1986)
for the stars listed in Table 5. For the purpose,
a 23 level model atom (22 levels for \OI\ plus the ground state of \OII ) with
43 bound-bound and 22 bound-free radiative transitions has been adopted
based on a model atom kindly provided by Dan Kiselman (2000, private communication).
The radiative data comes mainly from Opacity Project calculations (Seaton 1987).
As evidence is mounting that the simple-minded
recipe of Drawin (\cite{drawin68}) to incorporate the effects of inelastic collisions
with hydrogen overestimates the cross-sections by at least two orders of magnitude
(Lambert \cite{lambert93}; Belyayev et al. \cite{belyayev99}; Kiselman \cite{kiselman01};
Paul Barklem, 2001, private communication),
such collisions are not included in the calculations.
We have, however, performed test calculations with
H collisions estimated following Drawin's recipe multiplied with different
factors for all stars, as discussed briefly below.
Collisional excitation and ionization with electrons and charge transfer collisions
with hydrogen and protons from the ground levels of \OI\ and \OII\ are included. 
For details of the adopted atomic data we refer to Kiselman (1991, 1993)
and Carlsson \& Judge (\cite{carlsson93}).
We note that the non-LTE results for \OI\ are quite insensitive to including
additional levels and transitions as the \OI\ triplet line formation is well
described by a two-level atom line source function with the line opacity remaining
close to its LTE value (Kiselman \cite{kiselman93}).
The important factors for the line formation instead is the
line itself: the mean intensity dropping below the Planck function in the infrared,
the collisional cross-sections and the line strength.

The 1D LTE abundances for the three \OI\ triplet lines have been individually
corrected for the predicted departures from LTE and the resulting mean non-LTE 
oxygen abundance and the corresponding \ofe\ value are
given in the last two columns of Table 5. In all cases the non-LTE
effects have been estimated using MARCS model atmospheres
(Asplund et al. \cite{asplund97})
with the stellar parameters listed in Table 2. The non-LTE
abundance corrections amount to a 0.1-0.25 dex decrease of the LTE results,
in agreement with previous calculations without large H collisional
cross-sections (e.g. Kiselman 1991, 1993).
The results remain largely intact also when including the effects of
H collisions. With a factor of 0.01 times the Drawin estimates, the non-LTE
abundance corrections are within 0.01 dex of the case without H collisions.
Even when applying the Drawin values directly, the corrections only change
by $<0.04$ dex with the exception of \object{LP\,815-43} where the difference amount to
0.1 dex. We note that \object{LP\,815-43} has the largest non-LTE effect ($-0.25$ dex),
which becomes even larger at yet smaller metallicities.

\subsection{The influence of stellar granulation on \forOI\ and \FeII\ lines}

Although the \forOI\ and \FeII\ lines are virtually immune to departures
from LTE, the strengths of the lines depend on the detailed
structure of the adopted atmospheric model. Properties of particular importance
are the temperature gradient, the ratio of the line and continuous opacity
and the atmospheric inhomogeneities and velocity fields introduced by 
convection in the surface layers, also known as stellar granulation.
Since the \forOI\ and \FeII\ lines are sensitive to different parts of
the atmospheres due to the difference in excitation potential and ionization
balance, it can not
be taken for granted that the model atmosphere dependence cancels out
when forming \ofe\ abundance ratios, as is the case for
analyses using 1D model atmospheres.

In order to estimate the effects of stellar granulation on our derived O and
Fe abundances, the new generation of 3D, time-dependent hydrodynamical
model atmospheres (e.g. Asplund et al. \cite{asplund99},
2000a,b,c)
has been applied to study the line formation of the 
\forOI\ and \FeII\ lines. The procedure is the same as that adopted
for our recent investigation of the OH line formation 
(Asplund \& Garc\'{\i}a P{\'e}rez  \cite{asplund01}): a strictly differential comparison 
has been performed between 1D {\sc marcs} and 3D model atmospheres
with identical stellar parameters. For details of the convection simulations
the reader is referred to Asplund \& Garc\'{\i}a P{\'e}rez (\cite{asplund01}). 
The models employed here have a range of metallicities ($\feh = 0.0$ to $-2.0$
and correspond to solar-like stars ($\teff \simeq 5800$, $\logg =4.4$) 
and turn-off stars ($\teff \simeq 6200$, $\logg =4.0$). In addition
we have included the subgiant
\object{HD\,140283} ($\teff \simeq 5700$, $\logg =3.7$, $\feh = -2.5$) using
the 3D simulation presented in Asplund et al. (\cite{asplund99}).
The same spectrum synthesis and atomic data
(transition properties, continuous opacities, equation-of-state and
chemical composition) have been employed in both 1D and 3D to isolate the
impact of convection treatment from other possible sources of uncertainties.
For all line calculations the  assumption of LTE has been made,
which, as noted before, is well justified.
In the 3D case no macro- or microturbulence are needed 
while $\xi_{\rm micro} = 1.0$\,km\,s$^{-1}$ has been adopted for
the corresponding 1D calculations although only of some importance for the
\FeII\ lines in the solar metallicity models.

\begin{table}[t]
\caption{Comparison of the oxygen and iron LTE abundances derived with 1D 
hydrostatic and 3D hydrodynamical model atmospheres. 
The 3D abundances 
are those which reproduce the equivalent widths of the
\forOI\ and \FeII\ lines computed using a 
1D model atmosphere, a microturbulence of $\xi_{\rm micro} = 1.0$\,km\,s$^{-1}$ 
and the abundances given in the third and fourth columns.
}
\label{t:3D}
\begin{tabular}{lccccc} 
 \hline 
\noalign{\smallskip}
$T_{\rm eff}$ & log\,$g$ & [Fe/H] & [O/Fe]$^{\rm a}$ & 
$\Delta {\rm log} \epsilon {\rm (O)}$ &
$\Delta {\rm log} \epsilon {\rm (Fe)}^{\rm b}$ \\
 $$[K] & [cgs] & & &  
\smallskip \\
\hline 
\noalign{\smallskip}
5767 & 4.44 & $+0.0$ & $\pm0.00$ & $-0.04$ & $-0.02$ \\
     &      &        & $+0.20$ & $-0.04$ &  \\
     &      &        & $+0.40$ & $-0.03$ &  \\
5822 & 4.44 & $-1.0$ & $+0.20$ & $-0.08$ & $+0.04$ \\
     &      &        & $+0.40$ & $-0.08$ &  \\
     &      &        & $+0.60$ & $-0.06$ &  \\
5837 & 4.44 & $-2.0$ & $+0.40$ & $-0.12$ & $+0.10$ \\
     &      &        & $+0.60$ & $-0.12$ &  \\
     &      &        & $+0.80$ & $-0.12$ &  \\
6191 & 4.04 & $+0.0$ & $+0.00$ & $-0.01$ & $-0.01$ \\
     &      &        & $+0.20$ & $-0.01$ &  \\
     &      &        & $+0.40$ & $-0.01$ &  \\
6180 & 4.04 & $-1.0$ & $+0.20$ & $-0.07$ & $+0.02$ \\
     &      &        & $+0.40$ & $-0.11$ &  \\
     &      &        & $+0.60$ & $-0.11$ &  \\
6178 & 4.04 & $-2.0$ & $+0.40$ & $-0.16$ & $+0.06$ \\
     &      &        & $+0.60$ & $-0.17$ &  \\
     &      &        & $+0.80$ & $-0.15$ &  \\
5690 & 3.67 & $-2.5$ & $+0.40$ & $-0.22$ & $+0.05$ \\
     &      &        & $+0.60$ & $-0.24$ &  \\
     &      &        & $+0.80$ & $-0.24$ &  \\
\hline
\end{tabular}
\begin{list}{}{}
\item[$^{\rm a}$] For these calculations ${\rm log} \epsilon {\rm (O)} = 8.70$
and ${\rm log} \epsilon {\rm (Fe)} = 7.50$ have been adopted for the Sun.
\item[$^{\rm b}$] The impact of 3D model atmospheres on the derived
stellar metallicities have been investigated for in total 
six Fe\,{\sc ii} lines (6149, 6238, 6247, 6417, 6432 and 6456\,\AA). 
In all cases the different Fe\,{\sc ii} lines give the same granulation
corrections to within 0.03\,dex.
\end{list}
\end{table}

\begin{figure}
\resizebox{\hsize}{!}{\includegraphics{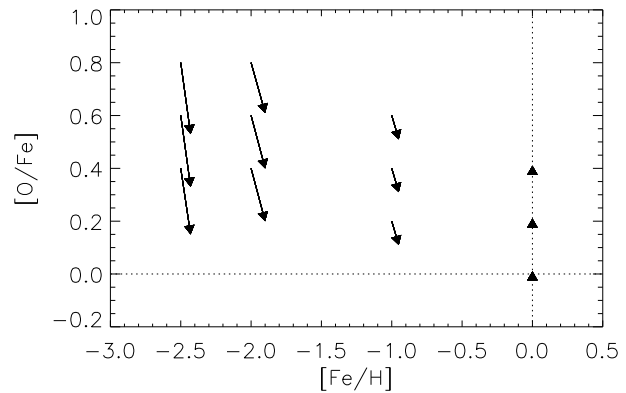}}
\caption{The effects of stellar granulation on \ofe\ based on \forOI\
and \FeII\ lines according to 3D hydrodynamical model atmospheres. Due
to the opposite behaviour of the \forOI\ and \FeII\ lines the impact 
on \ofe\ can be substantial at low metallicities. For these calculations
the adopted solar O and Fe abundances were 8.70 and 7.50, respectively.}
\label{f:OFe_3D}
\end{figure}

The results for the \forOI\ line and a subset of the \FeII\ lines utilized in
Sect. \ref{s:FeII} are summarized in Table \ref{t:3D}. 
The impact on \ofe\ is presented in Fig. \ref{f:OFe_3D} where the mean 
3D granulation corrections for the different \teff\ models are shown
relative to the solar case
\footnote{The results shown in Fig. \ref{f:OFe_3D} 
supersede those presented in Asplund (\cite{asplund_ga01})
 by the inclusion of \FeII\
lines, use of additional and more temporally
extended 3D model atmospheres and shown relative to the solar case.}. 
Even if the influence of granulation on the derived O and Fe abundances
are relatively small in most cases ($\simeq 0.1$\,dex), 
the combined effects on \ofe\ can
be substantial amounting to $\simeq -0.2$\,dex at $\feh\ = -2.0$
due to the opposite behaviour of the \forOI\ and \FeII\ lines.
In particular, the 3D correction to \ofe\ is $\simeq -0.26$\,dex for
\object{HD\,140283}.

\begin{figure}
\resizebox{\hsize}{!}{\includegraphics{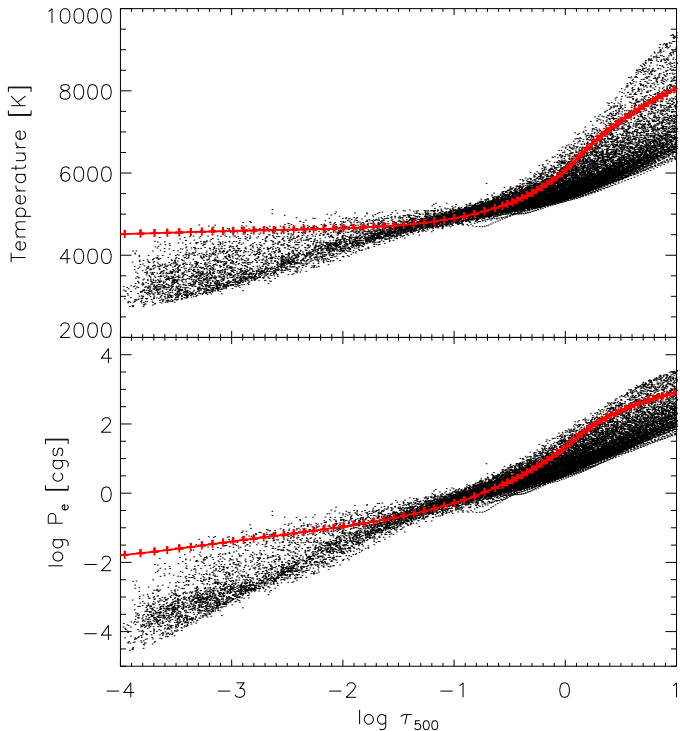}}
\caption{Temperature ({\em upper panel}) and electron pressure ({\em lower panel})
as a function of continuum optical depth at 500\,nm in one snapshot 
of the 3D hydrodynamical simulation of \object{HD\,140283}. Also
shown is the corresponding 1D {\sc marcs} photospheric structure 
(crosses connected with solid lines), illustrating the much lower temperatures
and consequently electron pressures encountered in 3D model atmospheres in
the upper atmospheric layers. }
\label{f:TPe_3D}
\end{figure}

The selected  \FeII\ lines  are predominantly  sensitive to 
the temperature gradient in
the deep atmospheric layers making them somewhat weaker
in 3D at low \feh\ (Asplund et al. \cite{asplund99}). The \forOI\
line on the other hand 
feels the cool, sub-radiative equilibrium upper layers at low metallicities,
making it significantly stronger in 3D than in 1D model atmospheres.
At first sight, this behaviour may seem peculiar since the population of 
the lower level of the \forOI\ line is essentially insensitive to the
temperature for these stars because it originates
from the ground state of the dominant ionization stage.
Instead the main effect on the line comes not from a change in line opacity
but from differences in continuous opacities between 3D and 1D models. 
The strengths of weak lines like the [O\,{\sc i}] (and Fe\,{\sc ii}) lines
 depend on 
$W_\lambda \propto \kappa_\nu^{\rm line}/\kappa_\nu^{\rm cont} \propto 
N_{\rm OI}(\chi = 0 {\rm ~eV})/N_{\rm HI} N_e \simeq
N_{\rm O}/N_{\rm H} \cdot 1/N_{\rm e}$
since the continuous opacity is provided mainly by H$^-$ around 630\,nm
and oxygen and hydrogen are predominantly in neutral form for these
atmospheres.
At low metallicities and the relevant temperatures, the number
density of
free electrons is very temperature sensitive. Thus, in order to compensate
for the lack of electrons in the cool surface layers in the 3D model
atmospheres, the abundance $N_{\rm O}/N_{\rm H}$ must be lowered to
retain the same equivalent width as in the corresponding 1D model
(Fig. \ref{f:TPe_3D}).

\begin{figure}
\resizebox{\hsize}{!}{\includegraphics{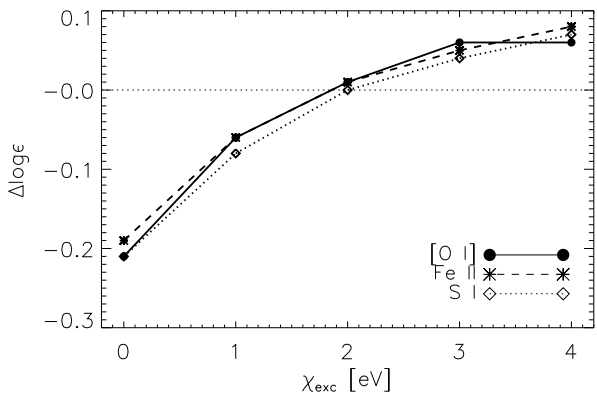}}
\caption{The 3D LTE abundance corrections (3D$-$1D) for a selection of fictitious
\OI , \FeII\ and \ion{S}{i} lines, all with an equivalent width of 0.5\,m\AA\
and located at 630\,nm but with different excitation potentials,
for the metal-poor subgiant {HD\,140283} ([Fe/H]$ =-2.5$). 
The lines of these majority species behave almost identically with low 
excitation potential lines being significantly stronger 
in 3D than in 1D model atmospheres.} 
\label{f:lines_3D}
\end{figure}

This effect is not restricted to the \forOI\ line but applies equally well
to any resonance line of a majority species, as illustrated in
Fig. \ref{f:lines_3D}. 
For this comparison a number of fictitious lines of \OI, \FeII\ and
\ion{S}{i} with different excitation potentials have been computed
for a few snapshots in the \object{HD\,140283} simulation as well as
for the corresponding 1D {\sc marcs} model atmosphere. The
$gf$-values for the different transitions have been adjusted to 
produce a line of similar strength as the real \forOI\ line, i.e.
$0.5$\,m\AA . 
In all cases, the lines of \OI, \FeII\ and \ion{S}{i} behave almost
identically since all are majority species. 
For low excitation potentials the effect on the continuous opacities
dominate while higher excitation lines, like the \FeII\ lines utilized
in the present study, are more sensitive to the effects on the line
opacities and thus the temperature gradient in the deeper atmospheric layers.

For the discussion below in terms of the true mean
trend in \ofe\ as a function of \feh\ it is reasonable to approximate 
the 3D results according to the simple relation
$\Delta \ofe \simeq 0.11 \cdot \feh$.
We emphasize that this relation can not be assumed to be automatically
valid for stellar parameters differing greatly from   those given in Table \ref{t:3D}.
In particular, giants and subgiants, 
like \object{BD\,$+23\degr3130$} which is discussed at some length below, 
may diverge from this relation given their lower \teff\ and \logg .
The sign is unlikely to change, however, and therefore \ofe\ ratios based
on \forOI\ and \FeII\ lines should always become smaller in 3D than in 1D.

\section{Discussion}

\subsection{\forOI\ and the oxygen problem}

\begin{figure*}
\sidecaption
\includegraphics[width=12cm]{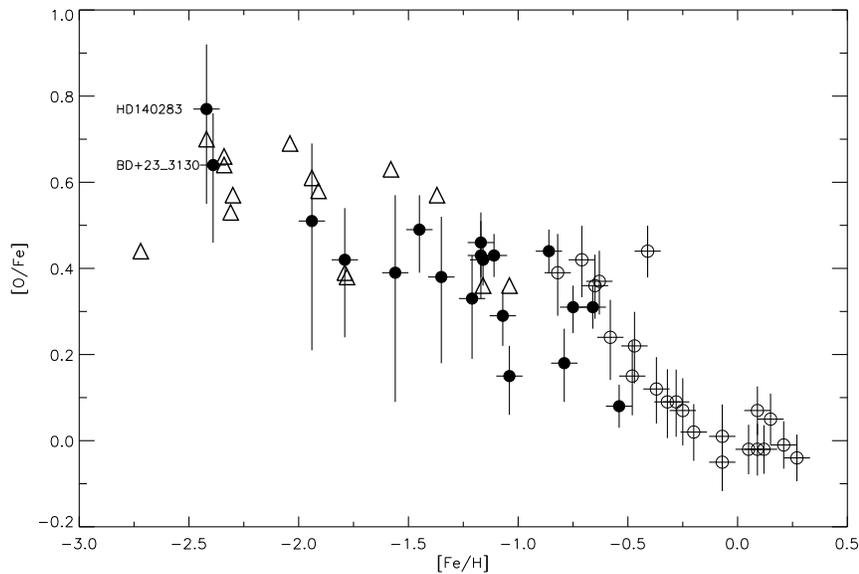}
\caption{\ofe\ vs. \feh\ as derived on the basis of 1D model atmospheres.
Filled circles refer to stars from the present paper and open circles
to disk stars from Nissen \& Edvardsson (\cite{nissen92}), in both cases
with the oxygen abundance determined from the [O\,{\sc i}] 6300.3 \AA\
line. The triangles refer to halo stars with non-LTE oxygen abundances
determined from the O\,{\sc i} triplet at 7774\,\AA .} 
\label{fig:ofe}
\end{figure*}

Our derived [O/Fe] values based upon the 1D hydrostatic model
atmospheres are plotted versus [Fe/H] in Fig. \ref{fig:ofe}
together with disk stars from Nissen \& Edvardsson (\cite{nissen92}).
Their \ofe\ values were
also obtained from the [O\,{\sc i}] 6300.3 \AA\ and Fe\,{\sc ii}
lines, but without any correction for the effect of the blending
Ni\,{\sc i} line. Hence, we calculated the equivalent width of the blend,
subtracted it from the observed equivalent width of the 6300.3 \AA\
feature, and recalculated the oxygen abundance. It is these new \ofe\
values for the disk stars that are plotted in Fig. \ref{fig:ofe}.
No attempt has been made to correct for the small offset between the
$T_{\rm eff}$ scale of Alonso et al. used here and that used by
Nissen \& Edvardsson, which was based on Edvardsson et al.'s
(\cite{edvardsson93})
theoretical calibration of the $b - y$ index. Although an offset
between the two sets of [O/Fe] values may be present, we expect it 
to be small due to the low sensitivity of [O/Fe] to $T_{\rm eff}$.

We emphasise again that the combination of [O\,{\sc i}] and
Fe\,{\sc ii} lines, as analysed with 1D model atmospheres yields
a result for [O/Fe] which is insensitive to reasonable errors in the
atmospheric parameters and to non-LTE effects. The [Fe/H] derived from
the Fe\,{\sc ii} lines is sensitive to the adopted surface gravity but
insensitive to non-LTE effects. Our surface gravity estimates are
not determined spectroscopically, as are many published values using 
the assumption of ionisation equilibrium. 
Although other spectroscopic indicators of the oxygen (and iron)
abundances may be affected by multiple sources of uncertainty,
the leading source of a systematic error in our abundance
determinations is likely to arise from
the simplifying assumptions entering into the calculation of
the 1D model atmospheres. 

As discussed in Sect. 4.4 and highlighted in Fig. \ref{f:OFe_3D}, the effect of
the stellar granulation as represented by the 3D hydrodynamic model
atmospheres is significant for the most metal-poor stars. While
emphasising that the grid of 3D models is as yet incomplete, we have
interpolated corrections from Table \ref{t:3D} and constructed Fig. \ref{fig:ofe3d} 
as an approximation to the [O/Fe] versus [Fe/H] for the 3D model
atmospheres. 

Our sample is the first using  accurate measurements of the [O\,{\sc i}]
line to provide extensive results
for a large number of metal-poor (say, [Fe/H] $\leq -1$) dwarfs and
subgiants. In light of the fact that published results using the
[O\,{\sc i}] line are
mostly based on inferior spectra, a detailed comparison is not attempted;
for example, Cavallo et al. (\cite{cavallo97}) draw from the
literature two quite different  measurements for the 6300 \AA\ line in \object{HD\,76932}
($W$ = 10 m\AA, and 1.8 m\AA) with neither one close to our measurement
of 3.3 m\AA .
Two metal-poor subgiants were analysed recently by
Fulbright \& Kraft (\cite{fulbright99}). 
Since one
-- \object{BD\,$+23\degr3130$} -- has been the object of several recent
discussions, we discuss it in some detail later
in this section.
Results from the [O\,{\sc i}] lines  in giants are discussed separately.

Attention was sharply focussed on oxygen abundances in metal-poor
stars by  studies of the OH ultraviolet lines (Israelian et
al. \cite{israelian98}, \cite{israelian01};
Boesgaard et al. \cite{boesgaard99}). These showed that
[O/Fe] indices derived from the OH
ultraviolet lines in metal-poor dwarfs were higher than the values
obtained previously from [O\,{\sc i}] lines in metal-poor giants.
The latter results had shown that [O/Fe] was
approximately constant ($\simeq 0.4$) for [Fe/H] $< -1$. In
sharp contrast, the OH-based results showed [O/Fe] to increase
quasi-linearly with decreasing [Fe/H].
The discrepancy between the abundance from OH in dwarfs and
[O\,{\sc i}] in giants was negligible  for [Fe/H] $\geq -1$ but increased
steadily to about 0.6 to 0.7 dex at [Fe/H] $\simeq -3$. 
Earlier, an analysis of the O\,{\sc i} triplet lines near 7774\,\AA\ in
metal-poor dwarfs had
also given high [O/Fe] values indicating a sharp disagreement with
results from the [O\,{\sc i}] 6300 \AA\ line in giants (Abia \&
Rebolo \cite{abia89}). 

With our analyses of the [O\,{\sc i}] line in dwarfs, we look 
afresh at the seemingly discrepant oxygen abundances. In
particular, we focus on three key   questions:

\begin{itemize}
\item
Do the [O\,{\sc i}] line and the OH 
lines  provide consistent oxygen abundances for  dwarfs and
subgiants?
\item
Do the [O\,{\sc i}] line and the O\,{\sc i} 7774 \AA\ triplet
provide similar abundances for dwarfs and subgiants?
\item
Is the oxygen abundance -- metallicity dependence similar for
dwarfs and giants?
\end{itemize}

\subsection{\forOI\  and {\rm OH} lines}

How do our new [O\,{\sc i}]-based abundances agree with those
derived from the OH ultraviolet lines, when both are obtained from
classical 1D model atmospheres?
A comparison is illuminating given the heat the `oxygen problem'
has generated in recent times.

Eight of our programme stars
were analysed by Israelian et al. (\cite{israelian98})
or Israelian et al. (\cite{israelian01}).
Differences in adopted model atmosphere parameters are small:
$\Delta T_{\rm eff} = 91 \pm$ 103 K, $\Delta \log$ g = 0.18 $\pm$
0.26, and $\Delta$[Fe/H] = 0.19 $\pm$ 0.07 where the sense of the
difference $\Delta$ is `this paper $-$ Israelian et al.'.
 Israelian et al.
chose their model atmosphere parameters and [Fe/H] from the literature.
Different grids of  1D model atmospheres were used in the two studies.

For four of the eight common stars, we determined the oxygen abundance 
from  the [O\,{\sc i}] line.
For this quartet which includes \object{HD\,140283} -- the most metal-poor
star in our sample -- the average difference between
 our abundances ([O/H] not [O/Fe]) from the
1D model atmospheres (Table 4)
and those from OH is  $-0.06$ dex.  If adjustments are made
for the use of  different grids of model
atmospheres and atmospheric parameters,  the mean difference may
increase  to about $-0.10$ dex for these stars with [O/H] between $-0.4$
and $-1.6$. Israelian et al. (\cite{israelian98})
adjusted the $gf$-values of the
OH lines to match the solar OH lines. In analysing  spectra of the
metal-poor stars, it seems preferable to adopt not solar $gf$-values
but the reliable theoretical-experimental $gf$-values. Mel\'{e}ndez et al.
(\cite{melendez01}) argue that were the latter $gf$-values adopted the
published OH-based abundances would be lowered by 0.1 - 0.2 dex, and
then the mean difference would increase from $-0.1$ dex to between
0.0 and 0.1 dex.

Six of our stars are in the sample analysed by Boesgaard et al.
(\cite{boesgaard99}), who
presented OH-based results for two different $T_{\rm eff}$ scales. Their
atmospheric parameters based on
Carney's (\cite{carney83b}) scale are similar to ours:
$\Delta T_{\rm eff}$ = 15 $\pm$ 34 K, $\Delta \log g$ = 0.09 $\pm$ 0.20,
and $\Delta$ [Fe/H] = 0.12 $\pm$ 0.07.
Of these six, we determined oxygen abundances for two (including \object{HD\,140283}).
After small corrections are applied for differences in the adopted
model atmospheres, differences in [O/H] are +0.1 and $-0.1$ dex.
Since Boesgaard et al. also adjusted $gf$-values to fit solar lines,
there will be, as in the case of Israelian et al., a slight revision
downward of the published abundances when laboratory $gf$-values are used.
\footnote{The O abundance derived from OH UV and IR lines is dependent on the
assumed dissociation energy of OH. Most analyses appear to adopt
$D_0 = 4.392$ eV, a recommendation by Huber \& Herzberg (\cite{huber79}) based on a
spectroscopic estimate by Carlone \& Dalby (\cite{carlone69}). A recent recommendation
provides $D_0 = 4.413 \pm 0.003$ eV (Ruscic et al. \cite{ruscic02}) from new experimental
and theoretical evidence and a reevaluation of the spectroscopic
data. Use of the new $D_0$ will result in slightly lower O abundances.
This revision is ignored here.}

These two comparisons show that, where a direct comparison is possible,
analyses of the ultraviolet OH and the 6300 \AA\ [O\,{\sc i}]
lines using 1-D model atmospheres give  consistent 
oxygen abundances for
metal-poor dwarfs and subgiants. This result applies to stars with 
[Fe/H] $> -2.5$.

Comparison of results is more commonly made in terms of the
run of [O/Fe] versus [Fe/H], but this introduces other considerations. 
The adopted solar O and Fe abundances now enter the
picture. Our adopted solar oxygen abundance is about 0.2 dex smaller than
chosen in the papers covering OH. Since the OH analyses used solar
$gf$-values, the lower solar abundance should not affect their
[O/Fe] values but it likely accounts for the fact that the solar $gf$-values
for OH lines 
were 0.1 - 0.2 dex smaller than the laboratory $gf$-values (see
above).
Additional effects should be noted. 
Kurucz models were used for the OH
analyses but MARCS models for the [O\,{\sc i}] line, but the effect on
[O/Fe] indices is likely small ($<$ 0.1 dex).
Also, our [Fe/H] are systematically slightly larger than those
adopted in the OH analyses.
The net effect of these various factors may not exceed about 0.1 dex in [O/Fe].

The OH-based results imply  a linear relation between
[O/Fe] and [Fe/H]
with a slope of $-0.33 \pm$ 0.02 (Israelian et al. \cite{israelian01}), a relation
which runs along  the upper boundary of the points in Fig. \ref{fig:ofe} 
for [Fe/H] $< -1.5$.
In short,  the two indicators give reasonably consistent results.

The much discussed subgiant star
\object{BD\,$+23\degr3130$}, which has been added to Fig. \ref{fig:ofe},
was thrust into the oxygen debate
by Fulbright \& Kraft (\cite{fulbright99}) who measured
the [O\,{\sc i}] 6300 \AA\
line and argued that the resulting [O/Fe] did not support the
high values from OH ultraviolet lines.
Others have discussed the [O\,{\sc i}] line in BD\,$+23\degr3130$
obtaining different results.
We have
reanalyzed \object{BD\,$+23\degr3130$} in the same way as our `own'
stars. We take the equivalent width of the 6300 \AA\ line
from Cayrel et al. (\cite{cayrel01}) who obtained
an UVES spectrum of \object{BD\,$+23\degr3130$} and determined the
equivalent width of \forOI\  6300.3~\AA\ to $W = 1.5 \pm 0.5$\,m\AA,
where the quoted $1\sigma$ error seems to be a rather conservative estimate
in view of the published high S/N spectrum.

From unpublished Str\"{o}mgren photometry of \object{BD\,$+23\degr3130$}
(Schuster et al.  \cite{schuster02}) $\teff (b-y) = 5195$\,K is
derived, and $V-K$ = 1.975
(Laird et al. \cite{laird88}) leads to $\teff (V-K) = 5145$\,K. In
both cases the IRFM \teff\ calibration for giants by Alonso et al.
(\cite{alonso99}) has been used and it was assumed that the
star is unreddened. Adopting the average \teff\ = 5170\,K,
the gravity derived via the Hipparcos parallax is $\logg = 3.0 \pm 0.25$.
Two of the \FeII\ lines from the list in Table 3,  5197.58 \AA\ and
5234.63 \AA, were observed by Fulbright \& Kraft (\cite{fulbright99})
and have equivalent widths of 19.6 and 23.0\,m\AA , respectively.
The corresponding metal abundance is $\feh = -2.39$. With these
parameters and using the equivalent width of the \forOI\  6300.3~\AA\
line given by Cayrel et al., we get  $\ofe = 0.64 \pm 0.15$, the value
plotted in Fig. \ref{fig:ofe}.
However, it cannot be excluded that \object{BD\,$+23\degr3130$} (distance $\simeq
230$ pc) is significantly reddened. Actually, the observed $H_{\beta}$
index of the star (Schuster et al. \cite{schuster02}) suggests
$\teff \simeq 5270$\,K
based again on the calibration of Alonso et al. (\cite{alonso99}).
This temperature would lead to $\ofe = 0.69 \pm 0.15$.

This shows that \object{BD\,$+23\degr3130$} confirms our result
for \object{HD\,140283}. 
The oxygen abundance for \object{BD\,$+23\degr3130$} is $\log\epsilon$(O)
= 6.99. Israelian et al. (2001) for a Kurucz model having almost
identical atmospheric parameters obtained $\log\epsilon$(O) = 7.10
from OH ultraviolet lines.
% If a MARCS model had been used, the OH-based abundance
%would be reduced slightly.
Balachandran et al.  (2001) also
adopted Cayrel et al.'s $W$ for the 6300 \AA\ line and with
atmospheric parameters essentially identical to ours  found
$\log\epsilon$(O) = 7.03 $\pm$ 0.15.
Reanalysis of Israelian et al.'s (\cite{israelian98})
equivalent widths of  OH ultraviolet lines with laboratory $gf$-values gave
$\log\epsilon$(O) = 7.06 $\pm$ 0.11, and analysis of their own 
measurements of OH infrared vibration-rotation lines gave
$\log\epsilon$(O) =  7.00 $\pm$ 0.07. Thus, 
as for other metal-poor stars,  
the [O\,{\sc i}] and OH ultraviolet (and infrared) lines
give concordant results. 

\begin{figure*}
\sidecaption
\includegraphics[width=12cm]{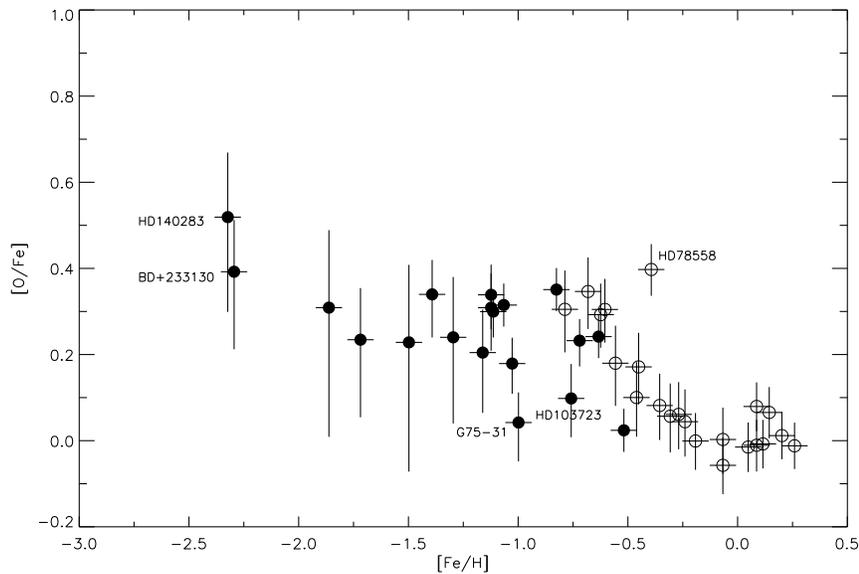}
\caption{\ofe\ vs. \feh\ after approximate correction for the effects
of stellar granulation.
Filled circles refer to stars from the present paper and open circles
to disk stars from Nissen \& Edvardsson (\cite{nissen92}), in both cases
with the oxygen abundance determined from the [O\,{\sc i}] 6300.3 \AA\
line.}
\label{fig:ofe3d}
\end{figure*}

The OH molecule also provides vibration-rotation lines from the ground
electronic state.  Fundamental or first overtone lines  have been
detected and analysed in a few cool metal-poor dwarfs.  Given spectra of
adequate $S/N$, the infrared lines have the advantages over the ultraviolet
lines that they lie in
a cleaner spectral interval, and are  likely to be formed
in LTE. 
In  addition to \object{BD\,$+23\degr3130$}
mentioned above, results are available for three additional metal-poor
dwarfs. Balachandran \& Carney (\cite{balachandran96}) discuss
\object{HD\,103095} (\object{Gmb\,1830}) obtaining
$\log\epsilon$(O) = 7.83 from OH infrared lines or [O/Fe] = 0.31  for
[Fe/H] = $-1.22$.
Mel\'{e}ndez et al. (\cite{melendez01}) give a similar result:
after correction for our lower solar oxygen abundance,  [O/Fe] $\simeq$
0.4 at [Fe/H] = $-1.36$.
These values fit the results in Fig. \ref{fig:ofe} for [Fe/H] $\simeq$
$-1.3$.  Results for two other dwarfs at [Fe/H] = $-0.8$ and $-1.8$  according
to  Mel\'{e}ndez et al. fall at the upper end of the distribution in
Fig. \ref{fig:ofe}. 
Balachandran et al.'s run of [O/Fe] versus [Fe/H]
from OH infrared lines in nine stars with [Fe/H] from $-1$ to $-3$ resembles
the distribution of points in Fig. \ref{fig:ofe}, but a 0.1 dex increase
for the lower solar oxygen abundance should
be applied.
A closer comparison is
not possible because details of the infrared analyses are unpublished.

The preceding comparisons between \forOI\ and OH lines were based on 
the use of 1D model atmospheres and ignored the effects
of granulation.
In Section 4.3, we discussed these effects  on the
\forOI\ line. Abundances corrected approximately for these
3D effects are shown in Fig. \ref{fig:ofe3d}.
Analogous effects on the 
OH lines were discussed by Asplund \& Garc\'{\i}a P\'{e}rez
(\cite{asplund01}). 
For a given effective temperature, the infrared and ultraviolet OH lines
are similarly sensitive to the effects of granulation (Asplund \&
Garc\'{\i}a P\'{e}rez 2001).
The effects of granulation on the OH lines appears to decline with
decreasing effective temperature.
Since stars observed for their infrared OH
lines are generally cooler than those observed for the ultraviolet
lines, the abundance corrections applicable in the case of the infrared OH lines
may in practice be smaller than for the ultraviolet lines
(with the exception that not yet investigated non-LTE effects may impact
the UV but not the IR lines).
We note though that no 3D models with parameters appropriate for the
IR targets at low metallicities are yet available to verify the
continuation of the 3D behaviour down to these lower \teff .

The 3D corrections of abundances derived from \forOI\ and OH are qualitatively
similar in that the corrections increase with decreasing metallicity,
but the corrections for OH are  larger than for [O\,{\sc i}]. 
At [Fe/H] = $-2.5$, the approximate limit of our sample, the OH-based
O abundance is reduced by about 0.5 dex at \teff\ = 5800 K,
and by about 0.7 dex at \teff\ = 6200 K, but
the [O\,{\sc i}]-based abundances are decreased by only about 0.2 dex.
Given that we have found OH and \forOI\ lines to give
essentially the same abundance with the 1D atmospheres, an
application of the 3D atmospheres would apparently reopen the
oxygen problem that we have just closed. A resolution of this
problem may lie in the replacement of the assumption of
LTE by non-LTE. The \forOI\ line is immune to non-LTE effects
but formation of OH ultraviolet lines in non-LTE  will lead to 
weaker lines, and, hence, the need for a higher O
abundance; Asplund \& Garc\'{\i}a P\'{e}rez estimate a rough
correction of 0.2 dex but with little [Fe/H] sensitivity.
This correction would not apply to the OH infrared lines. 
An additional factor applicable to both ultraviolet and
infrared lines is that the OH density may not be at its
LTE value throughout the atmosphere when granular flows
are present. 

Since the \forOI\ and \FeII\ lines are immune to non-LTE
effects and the corrections for stellar granulation are relative
small, we present our abundances with the estimated
corrections from the 3D model atmospheres as the best
indicator of oxygen abundances for metal-poor dwarfs (Fig. \ref{fig:ofe3d}).
After correction for stellar granulation, the index [O/Fe] 
rises with decreasing [Fe/H] to reach the `low' value
[O/Fe] $\simeq$ 0.3 - 0.4 for [Fe/H] from about $-0.7$ to $-2.0$
with perhaps an increase to lower  [Fe/H].
Certainly, the OH ultraviolet lines as analysed using 1D atmospheres
indicate a rising [O/Fe], but the 3D atmospheres suggest an
increasing correction with decreasing [Fe/H] which cancels
much of the apparent rise.

By showing that the OH and [O\,{\sc i}] lines give consistent
abundances when analysed with 1D atmospheres, we appear to have
removed a key part of the oxygen problem. Yet, the 3D atmospheres
by suggesting lower oxygen abundances from OH lines than from the
[O\,{\sc i}] line introduce an inverse of the original
problem. While consideration of non-LTE effects may yet
correct this inverse problem, 
 the largest
potential source of  systematic errors in oxygen
abundance determinations may reside in 
present realisations of stellar granulation.  

\subsection{\forOI\ and {\rm O\,{\sc i}} lines}

The permitted triplet of \OI\ lines near 7774\,\AA\ has been
analysed extensively in metal-poor dwarfs and giants.
Analysts of the OH ultraviolet lines have argued that
the OH and O\,{\sc i} lines provide consistent oxygen abundances.
Other 
analyses of the O\,{\sc i} lines have
confirmed that the O\,{\sc i} lines give `high' [O/Fe]
values for metal-poor stars
(e.g., Cavallo et al. \cite{cavallo97}; Mishenina et al.
\cite{mishenina00}).

As described in Sect. 4.3 we determined oxygen abundances from a non-LTE
analysis of the \OI\ triplet for 15 stars. The resulting 
\ofe\ values are plotted in Fig. \ref{fig:ofe}. As seen
the triplet gives [O/Fe] indices similar to the 1D values
from  the \forOI\ line. For five of the 15 stars, we
used both the permitted and the forbidden line finding differences
(permitted - forbidden)
ranging from $-0.07$ to 0.21 dex and a mean difference of 0.03 dex.
This good agreement exists for analyses using the 1D model atmospheres.

Present indications (Asplund \cite{asplund_ga01}) are that a LTE
analysis using an equivalent 3D model atmosphere leads to a
less than 0.1 dex  change in
the oxygen abundance from O\,{\sc i} lines.
Then, application of 3D model atmospheres reduces the abundance
from the \forOI\ line more than that from the \OI\
lines. Unless the non-LTE corrections to the O\,{\sc i}
lines are enhanced in the 3D model atmospheres,  
oxygen abundances from the O\,{\sc i} lines may be
$\simeq 0.2$~dex larger than from the [O\,{\sc i}] line. 

\subsection{Oxygen abundances in dwarfs and giants}

Arguments favoring the [O\,{\sc i}] - Fe\,{\sc ii} combination of
lines as the optimal indicator of  the [O/Fe] ratio hold for
giants as well as dwarfs. Indeed, the run of [O/Fe] versus [Fe/H]
was considered as well established from [O\,{\sc i}] lines 
in giants well before the appearance of the 
measurements of OH ultraviolet lines in dwarfs.
For example, Barbuy (\cite{barbuy88})
showed that [O/Fe] $\simeq$ +0.4 from analyses of  metal-poor
giants with [Fe/H] extending down to $-2.5$.  (Her estimates of [Fe/H]
are not based on Fe\,{\sc ii} but Fe\,{\sc i} lines,
although the few illustrated fits to observed spectra imply a good fit to
the Sc\,{\sc ii} line near 6300.6 \AA). Other  analyses
using the [O\,{\sc i}] line gave similar results for [O/Fe]
(e.g., Lambert et al. \cite{lambert74};
Gratton \& Ortolani \cite{gratton86};
Sneden et al. \cite{sneden91}; Kraft et al. \cite{kraft92}).
Since most analyses referenced the stellar oxygen abundance
to a solar abundance approximately 0.2 dex higher than our
adopted value, and used a $\log gf$ value 0.03 dex smaller than
our value, the published [O/Fe] should be increased by
about 0.2 dex. In some cases, a revision should be applied also
to account for different values of the solar iron abundance. Early
analyses may have adopted a solar abundance $\log\epsilon$(Fe) = 7.65
instead of the now favored value close to $\log\epsilon$(Fe) = 7.50.
Such revisions simultaneously to the solar O and Fe abundances cancel
almost exactly for [O/Fe]. 

Recently, Sneden \&
Primas (\cite{sneden01})  obtained high-quality spectra and analysed 
the [O\,{\sc i}] 6300 \AA\ line alongwith
Sc\,{\sc ii} and Fe\,{\sc ii} lines to give
[O/Fe] and [O/Sc] with [Sc/Fe] $\simeq$ 0.0  expected from many
studies.  After adjustment for our lower solar oxygen
abundance and higher $gf$-value for the forbidden line,
their preliminary result 
is [O/Fe] $\simeq$ +0.45 with an indication of a slight
increase with decreasing [Fe/H] to reach [O/Fe] $\simeq$ +0.6 at
[Fe/H] = $-2.7$ from [O/Fe] $\simeq$ +0.45 at [Fe/H] = $-1.2$.
In view of the
insensitivity of the [O/Fe] results to the adopted model atmospheres,
these results are unlikely to change appreciably for alternative
choices of effective temperature, and  surface gravity.

King (\cite{king00})  discussed published analyses of [O\,{\sc i}] and
Fe\,{\sc ii} lines in metal-poor giants and obtained 
almost identical results (see his Fig. 5). Much of King's
paper is concerned with the application of non-LTE corrections
to published iron abundances derived from Fe\,{\sc i}
as a way to reconcile disagreements
over the run of [O/Fe] with [Fe/H].  By focussing on [O\,{\sc i}]
and Fe\,{\sc ii} lines, non-LTE becomes a moot issue.  Insensitivity
to non-LTE effects is a powerful advantage of using Fe\,{\sc ii}
lines given the preliminary nature of extant non-LTE predictions
for iron.

At [O/Fe] $\simeq$ +0.5, the giants'  [O\,{\sc i}] and
Fe\,{\sc ii} lines provide results for [O/Fe] comparable to those in
Fig. \ref{fig:ofe}.
The predicted effects of granulation
on [O/Fe] for dwarfs  lower the ratio significantly from
values obtained using 1D model atmospheres. Preliminary corrections
give the results shown in Fig. \ref{fig:ofe3d}, where [O/Fe] $\simeq$ +0.3 for
$-2.0 < \feh < -1.0$ for dwarfs.
It is these results that are in mild disagreement
with those from giants obtained from 
the [O\,{\sc i}] -- Fe\,{\sc ii} combination
of lines and 1D model atmospheres.  
To take the final step in the application of the [O\,{\sc i}] and
Fe\,{\sc ii} lines to a definitive determination of  [O/Fe]
for metal-poor stars, it is necessary to  apply
3D model atmospheres to analyses of lines in spectra of giants, but
such
3D model atmospheres do not yet exist. 
One might speculate  that effects of stellar granulation on the
lines in giants will lower [O/Fe] by about 0.2 dex. 

The OH vibration-rotation lines but not the ultraviolet lines have
been used to obtain oxygen abundances for giants (Mel\'{e}ndez et al.
\cite{melendez01}). The published results depend on the choice of atmospheric
parameters: [O/Fe] = +0.5 or +0.3. These values should be increased
by about 0.13 dex to account for the reduction in the solar oxygen
abundance. A mean of the two values with the increase of 0.13 dex gives
[O/Fe] $\simeq$ 0.5, a result consistent with that from the
forbidden oxygen line.

Oxygen triplet lines in giants ($\log g$ $<$ 3.0) were observed and
analysed by Cavallo et al. (\cite{cavallo97}). Their LTE results for [O/Fe] for
[Fe/H] $< -1$ fall  above our points in Fig. \ref{fig:ofe}: [O/Fe] $\simeq$
+0.8 for [Fe/H] $< -1$.  Mishenina
et al. (\cite{mishenina00}) observed a few additional giants and undertook a
non-LTE analysis for their stars as well as Cavallo et al.'s
sample. Non-LTE effects as implemented by Mishenina et al.
reduce [O/Fe] by about 0.1 dex on average, and the
non-LTE oxygen abundances leave [O/Fe] 
significantly above ours in Fig. \ref{fig:ofe}:
[O/Fe] $\simeq$ 0.37[Fe/H] + 0.11
is a fit by eye to Mishenina et al.'s results for giants. 
Our lower solar oxygen abundance implies that the published
[O/Fe] should be increased by 0.2 dex.
 A critical analysis of the atmospheric parameters
for giants including the [Fe/H] indices might be instructive;
we note that for three of the four stars in common between
Cavallo et al. and Johnson \& Bolte (\cite{johnson01}), an example of a
recent spectroscopic analysis, there are large differences in
the adopted atmospheric parameters with the recent choices leading
apparently to larger [O/Fe] indices.  It remains to consider  the
non-LTE effects on the iron lines, or to minimize their influence by
analysing the Fe\,{\sc ii} lines. In the context of 1D model
atmospheres, the high oxygen abundance for giants from the triplet lines
remains the outstanding issue of the oxygen problem.

\section{Concluding remarks}

In this paper, we have demonstrated that an efficient high-resolution
spectrograph on a very large telescope can now provide firm detections of
the weak [O\,{\sc i}] lines 
in spectra of metal-poor dwarfs. Analysis of the forbidden oxygen
lines with Fe\,{\sc ii} lines provides an estimate of [O/Fe] which is
insensitive to the adopted model atmosphere parameters \teff\ and \logg .

Our results using standard 1D model atmospheres show
[O/Fe] to increase with decreasing [Fe/H] (Fig. \ref{fig:ofe}). 
We have shown that the oxygen problem largely vanishes when 1D models are
applied; in particular
we find a satisfactory agreement between oxygen abundances
derived from the [O\,{\sc i}] 6300 \AA\ line in dwarfs and subgiants  
and abundances obtained from OH ultraviolet and infrared lines, and the
O\,{\sc i} triplet lines. Published results for [O/Fe] in 
metal-poor giants using also the [O\,{\sc i}] -- Fe\,{\sc ii} combination
of lines are compatible with results in Fig. \ref{fig:ofe}
from the same combination.
Observations of the O\,{\sc i} lines in
giants give LTE abundances and [O/Fe] values larger than our
[O\,{\sc i}]-based results in Fig. \ref{fig:ofe}. Non-LTE effects reduce the
discrepancy. 

Application of 3D model atmospheres reintroduces in a mild fashion
aspects of the oxygen problem.
The LTE OH-based oxygen abundances for dwarfs appear
to be less than the [O\,{\sc i}]-based results, but the O\,{\sc i}-based
results are appreciably larger than the   [O\,{\sc i}]-based
results in Fig. \ref{fig:ofe3d}. Non-LTE effects yet to be quantitatively
assessed for OH and O\,{\sc i} lines in the 3D atmospheres may
reduce these differences relative to the non-LTE insensitive
[O\,{\sc i}] line. There are also small abundance
differences between results from 3D atmospheres for dwarfs
and 1D atmospheres for giants. Differences arising from the introduction of
the 3D atmospheres should challenge the confidence of stellar
spectroscopists using the now extensive grids of 1D model
atmospheres. To respond to  this challenge
theoretical studies of atmospheric inhomogeneities (granulation)
must be pursued. There is an urgent need for 3D model
atmospheres for giants. A parallel effort on observational
quantification of the effects of granulation on spectral
lines (and the continuum fluxes)  must be carried out: measurements of
line asymmetries and velocity shifts, as well as  equivalent
widths  of lines. Lack of understanding of stellar granulation
may presently be the primary source of systematic errors in
abundance analyses. Evidence presented here indicates that
the magnitude of these errors may well exceed the errors
often confidently attached to abundances derived from classical
model atmospheres. 

Finally, we note that in describing plots like Fig. \ref{fig:ofe3d}
there is a tendency to
impose a simple linear relation onto the points. Instead one might
identify a plateau at [O/Fe] $\simeq$ +0.3 for [Fe/H] from about
$-0.7$ to $-2.0$ with a rise to higher [O/Fe] at lower [Fe/H].
Furthermore, not all stars may lie on a single relation.
Although much of the apparent scatter about a
mean relation is attributable to measurement errors, there
are a few stars that most probably fall off a mean relation. \object{HD\,103273}
with a lower than average oxygen abundance is a halo star
known to be unusually deficient in Mg, Si, and Ca
($\alpha$-elements) (Nissen \& Schuster \cite{nissen97}), and a
slight oxygen deficiency fits this pattern. \object{G\,75-31} may be another
example of an $\alpha$-poor star. \object{HD\,78558} is a thick-disk star
with an  above average [$\alpha$/Fe] and, hence, its
unusually high [O/Fe] is not unexpected. The star has
a small mean galactocentric distance according to Nissen \& Edvardsson
(\cite{nissen92}). In order to investigate the reality of this possible
cosmic scatter of \ofe\ at a given value of \feh\ a much larger sample
spanning stars with different Galactic orbits should be studied.

\begin{acknowledgements}
Drs. A. Alonso and G. Simon are thanked for sending $K$ magnitudes
of some stars in advance of publication.
The authors are indebted to Dan Kiselman for kindly sharing his expertise
in non-LTE line formation and his \OI\ model atoms.
MA and PEN acknowledge the Swedish and  Danish Natural Science Research
Councils for financial support of the present project.
DLL thanks the R. A. Welch Foundation of Houston, Texas (grant F-634)
for support.  This research has made use of
the SIMBAD database operated at CDS, Strasbourg, France.
\end{acknowledgements}

{}

\end{document}